\documentclass [11 pt, twoside, a4paper]{article}

\usepackage[utf8]{inputenc}

\usepackage[autostyle]{csquotes}
\usepackage[english]{babel}
\usepackage[T1]{fontenc}

\usepackage[sorting=none, bibstyle=numeric,backend=biber]{biblatex}
\usepackage{makecell}
\usepackage{amsfonts}
\usepackage{graphicx}
\usepackage{tabularx}
\usepackage{adjustbox}
\usepackage{amsmath}
\usepackage{amsthm}
\usepackage{cases}
\usepackage{epigraph}
\usepackage{amssymb} 
\usepackage{hyperref}
\usepackage{epstopdf}
\usepackage{latexsym}
\usepackage{eucal}
\usepackage{float}
\usepackage{verbatim}
\usepackage{fancyhdr}
\usepackage{setspace}
\usepackage{graphicx}
\usepackage{url}
\usepackage{bm}
\usepackage{siunitx}
\usepackage{caption}
\usepackage{subfig}
\DeclareCaptionLabelFormat{onlynumber}{Figure (\Alph)}
\DeclareCaptionLabelFormat{Anumber}{A#2}
\DeclareCaptionLabelFormat{3number}{3#2}
\DeclareCaptionLabelFormat{Hnumber}{H#2}

\newcolumntype{L}{>{\raggedright\arraybackslash}X}
\DeclareUnicodeCharacter{0301}{\'{e}}

\fancyhfoffset[L]{0.7cm} % left extra length
\fancyhfoffset[R]{0.7cm} % right extra length
\theoremstyle{plain}

\theoremstyle{definition}

\theoremstyle{remark}

\usepackage[paper=a4paper,margin=1in]{geometry}

\usepackage{listings}
\usepackage{color}

\definecolor{dkgreen}{rgb}{0,0.6,0}
\definecolor{gray}{rgb}{0.5,0.5,0.5}
\definecolor{mauve}{rgb}{0.58,0,0.82}

\usepackage{nameref}
\usepackage{varioref}
\usepackage{hyperref}
\usepackage{cleveref}

\begin{document}
\title{\MakeUppercase{\textbf{Pseudo-online framework for BCI evaluation: a MOABB perspective}}}
\author{Igor Carrara$^{1, 2}$ Th\'eodore Papadopoulo$^{1, 2}$ \\
\small $^1$ Université Côte d'Azur (UCA)\\
\small $^2$ Centre Inria d'Université Côte d'Azur, Cronos Team \\
\small igor.carrara@inria.fr and theodore.papadopoulo@inria.fr}
\date{}
% \author{First A. Author, \IEEEmembership{Fellow, IEEE}, Second B. Author, and Third C. Author, Jr., \IEEEmembership{Member, IEEE}

\maketitle

\begin{abstract}
% OBJECTIVE 300 words
\textit{Objective}: BCI (Brain-Computer Interface) technology operates in three modes: \textit{online}, \textit{offline}, and \textit{pseudo-online}. In the \textit{online} mode, real-time EEG data is constantly analyzed. In \textit{offline} mode, the signal is acquired and processed afterwards. The \textit{pseudo-online} mode processes collected data as if they were received in real-time. The main difference is that the \textit{offline} mode often analyzes the whole data, while the \textit{online} and \textit{pseudo-online} modes only analyze data in short time windows. \textit{Offline} analysis is usually done with asynchronous BCIs, which restricts analysis to predefined time windows. Asynchronous BCI, compatible with \textit{online} and \textit{pseudo-online} modes, allows flexible mental activity duration. \textit{Offline} processing tends to be more accurate, while \textit{online} analysis is better for therapeutic applications. \textit{Pseudo-online} implementation approximates \textit{online} processing without real-time constraints. Many BCI studies being \textit{offline} introduce biases compared to real-life scenarios, impacting classification algorithm performance. 
\textit{Approach}: The objective of this research paper is therefore to extend the current MOABB framework, operating in \textit{offline} mode, so as to allow a comparison of different algorithms in a \textit{pseudo-online} setting with the use of a technology based on overlapping sliding windows. To do this will require the introduction of a idle state event in the dataset that takes into account all different possibilities that are not task thinking. To validate the performance of the algorithms we will use the normalized Matthews Correlation Coefficient (nMCC) and the Information Transfer Rate (ITR). 
\textit{Main results}: We analyzed the state-of-the-art algorithms of the last 15 years over several Motor Imagery (MI)
datasets composed by several subjects, showing the differences between the two approaches from a statistical point of view. \textit{Significance}: The ability to analyze the performance of different algorithms in \textit{offline} and \textit{pseudo-online} modes will allow the BCI community to obtain more accurate and comprehensive reports regarding the performance of classification algorithms.

\end{abstract}

\paragraph{Keywords} BCI-EEG, Asynchronous BCI, Riemann Geometry, MOABB, Pseudo Online BCI, Deep Learning, Machine Learning. 

\section{Introduction}
\label{Introduction}
Brain Computer Interface (BCI) is a technology that allows a digital device to be controlled through brain activity signals.
In recent years, many diverse modalities for acquiring the signal produced by the brain during a specific cognitive task have been developed. In general, we can categorize such procedures into non invasive, with techniques like Electroencephalogram (EEG)~\cite{berger:29} or invasive as the recent Endovascular Electrodes~\cite{oxley-yoo-etal:21}.
EEG is a non-invasive acquisition technique, with high time resolution and is relatively inexpensive. For these reasons, we will focus on BCI-EEG.
During this research, we will focus on Motor Imagery (MI) tasks, i.e., when the user changes his mental activity by thinking of performing a particular body movement, but the overall framework is generic and can be applied in many different BCI contexts.

A BCI technology can operate in 3 different modalities: the \textit{online} mode, which requires to constantly analyze the new input data based on real-time EEG data, the \textit{offline} mode where the signal is first acquired and saved, and then processed later with no real time constraints. Lastly, the \textit{pseudo-online} mode does not process the data in real-time during the experiment but the collected data are processed a posteriori as if they were received online.
The main differences are that in the \textit{offline} mode the whole data is available for the analysis, while in the \textit{online} or \textit{pseudo-online} modes, the data is typically analyzed in a short time window running across the signal. The \textit{online} and \textit{pseudo-online} differ by the amount of time that can be used to process the data i.e. the \textit{pseudo-online} method analyzes the same data as the \textit{online} method but with no real time constraint on the processing time.

\textit{Offline} analysis is usually done with synchronous BCIs, i.e. BCIs that process the signal in predefined time windows where a mental task is performed (e.g. the imagination of a movement) and discards the remaining signal, thus creating a mode of interaction that is unnatural for real-life applications. In contrast, an asynchronous BCI, compatible both with \textit{online} and \textit{pseudo-online} modalities allows a given mental activity to be performed for the duration decided by the subject and not restricted to specific time windows. In particular, such BCIs must be able to distinguish the brain signal between intervals of rest or idle periods vs mental activity. It might as well be able to decide between different types of mental activities.

Usually \textit{offline} processing of the EEG signals turns out to be more accurate, while a \textit{online} signal analysis approach generally produces results that are less accurate but more better suited for use in a therapeutic application~\cite{rodriguez-ugarte-ianez-etal:17}.
The \textit{pseudo-online} implementation can be used as a methodology that best approximates the \textit{online} process while relaxing the real-time constraint for the processing, thus showing the best attainable performance for a specific BCI task.
Many BCI studies are tested only \textit{offline}, thus generating unrealistic performance compared to real-life scenarios~\cite{lehtonen-jylanki-etal:08}. 
Such approaches introduce an important research bias, as many new classification algorithms created to perform well \textit{offline} lose their competitive nature in real-life applications.
It is therefore of particular importance for the advancement of BCI technology that algorithms are validated in \textit{online} or \textit{pseudo-online} mode. Some studies test their algorithms \textit{online}, but their datasets and codes are not always made public, making the data analysis unreproducible. All of this has an extremely negative influence on the speed of progress in the BCI field, making it particularly difficult and complex for people to reproduce published results. As a matter of fact, even just trying to reproduce the performance of state-of-the-art algorithms on a specific dataset is complex and time demanding. In addition, the subjects collected in each dataset are usually few, which is statistically non significant, thus comparing different algorithms on different datasets can produce even antithetical results.

To solve some of these issues for the \textit{offline} mode, the MOABB~\cite{jayaram-barachant:18} framework was introduced to test the performance of different classification algorithms on identical datasets and identical preprocessing pipelines. This framework has been a turning point for the BCI community. However, it does not currently include \textit{pseudo-online} testing, thus having a lower impact on \textit{online} BCI quality. 

Our first contribution is to propose a \textit{pseudo-online} extension of MOABB, with the use of a technology based on overlapping sliding windows, which also enables the integration of analyses in asynchronous mode. 

In addition, we created a performance dashboard comparing the best-known algorithms in BCI classification. This dashboard can be used as a starting point for comparing one's own new algorithm.
We took care to test the best state-of-the-art pipelines produced in the last 15 years. This list cannot be understood as definitive but rather as a starting point.
To perform this comparison, it was also necessary to extend MOABB, currently based on the scikit-learn library, with the deep learning frameworks of TensorFlow and Keras.

Ultimately, the goal of this paper is to introduce a framework for \textit{pseudo online} analysis of BCI datasets, so as to enable rapid advancement of performance in the BCI community and also to make the community more inclusive to people with different backgrounds. 
The framework is showcased with EEG Motor Imagery with 4 datasets but the framework can be easily extended to other MI datasets, data acquisition procedures or even other types of BCIs. 

The following article is structured as follows: in \ref{Methods}, we describe the framework and the pipelines considered in the state of the art; \ref{Results} lists the results obtained using within-session and cross-session evaluation. Finally, \ref{Discussion} analyzes the implications of the framework and its current limitations; \ref{Conclusion} summarizes the results of our study.

\section{Methods}
\label{Methods}
In this section, we list the different Motor Imagery (MI) datasets considered in this research, along with the methodology  used to transform a regular \textit{offline} dataset into a format suitable for a \textit{pseudo online} analysis. This procedure is applicable to every dataset recorded using normal procedures to be processed \textit{offline}, providing a great versatility to the framework. This concept of extending a synchronously recorded dataset to an asynchronous approach has already been proposed in~\cite{sadeghian-moradi:07}.

\begin{figure}[!ht]
 \centering
 \includegraphics[width=\linewidth]{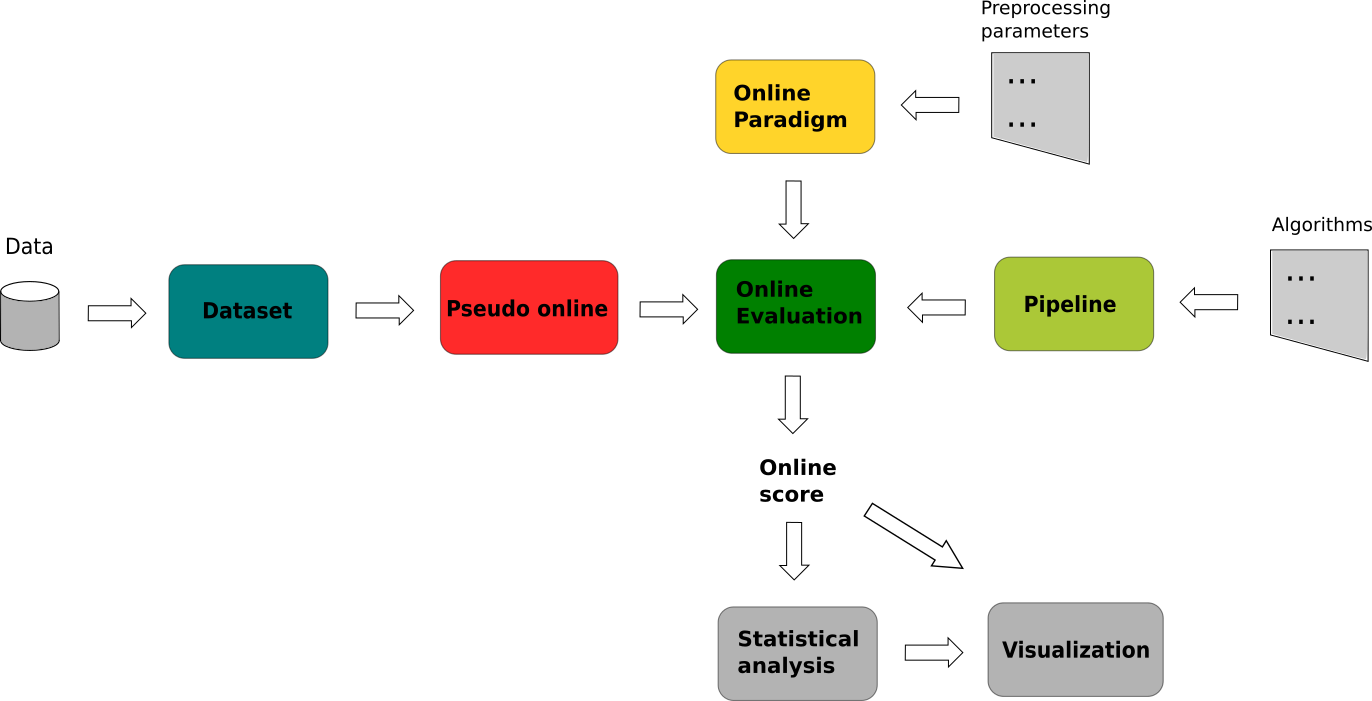}
 \caption{Representation of the framework for the \textit{pseudo-online} architecture, partially inspired by~\cite{jayaram-barachant:18}.}
  \label{fig:fig1}
\end{figure}

We will also explain the concept of paradigm and the possible different evaluation procedures. We will then present the statistical analysis that we used and devote special attention to the metrics considered in that analysis. The differences of the proposed framework with respect to the standard MOABB is described in the Figure~\ref{fig:fig1}.
The whole project is implemented using Python3 and is based on the use of the MNE~\cite{gramfort-luessi-etal:13}, PyRiemann~\cite{barachant-king:15}, scikit-learn~\cite{pedregosa-varoquaux-etal:11}, TensorFlow~\cite{abadi-agarwal-etal:15}, MOABB~\cite{jayaram-barachant:18} and SciKeras libraries.

\subsection{Datasets}
We consider 4 open-access motor imagery \textit{offline} BCI datasets consisting of several subjects for each dataset and several sessions for each subject.
\ref{table:dataset} contains all the details about these datasets.

\begin{table}[ht!]
\caption{Datasets considered during this study.}
\label{table:dataset}
\begin{center}
\begin{tabular}{c|c|c|c|c|c}
Dataset & Subjects & Channels & Sampling Rate & Sessions & Task \\
\hline BNCI2014001~\cite{tangermann-muller-etal:12} & 9 & 22 & 250 Hz & 2 & 4 \\
\hline BNCI2015001~\cite{faller-vidaurre-etal:12} & 12 & 13 & 512 Hz & 1 & 2 \\
\hline BNCI2014002~\cite{steyrl-scherer-etal:16} & 14 & 15 & 512 Hz & 5 & 2 \\
\hline BNCI2014004~\cite{leeb-lee-etal:07} & 9 & 3 & 250 Hz & 1 & 2 \\  
\hline 
\end{tabular}
\end{center}
\end{table}

Each of these datasets include a stimulus channel (\textit{stim} channel) that marks an events only when the subject is actively engaged in a task. 
To align the datasets with online situations, the first step to transform it is to introduce a \textit{nothing} event for each part that is not associated to a task. 
This inclusion allows for a performance evaluation that better reflects real-life scenarios. In practical applications, individuals may have periods where they actively attempt to perform a task, while they may be engaged in various unrelated thoughts such as daydreaming at other times. The \textit{nothing} event is designed to encapsulate these diverse possibilities that are not task-related.

The introduction of the \textit{nothing} event, however, introduces an important issue; the dataset now turns out to be strongly unbalanced toward that new class. We will analyze this problem and propose possible solutions in~\ref{metric}.

Using this procedure, we were able to test algorithms for the classification of 5 MI tasks (BNCI2014001) or 3 MI tasks (BNCI2014002, BNCI2014004, BNCI2015001). 

\subsection{Paradigm}
Following the line drawn by MOABB, we consider the paradigm as a way to transform continuous data to trials, i.e., the basic elements for any machine learning algorithm. 
In addition, the paradigm is used to set the preprocessing of the continuous data, keeping it unique for all datasets and all subjects considered in order to allow a fair comparison.

\begin{figure}[!ht]
 \centering
 \includegraphics[width=0.7\linewidth]{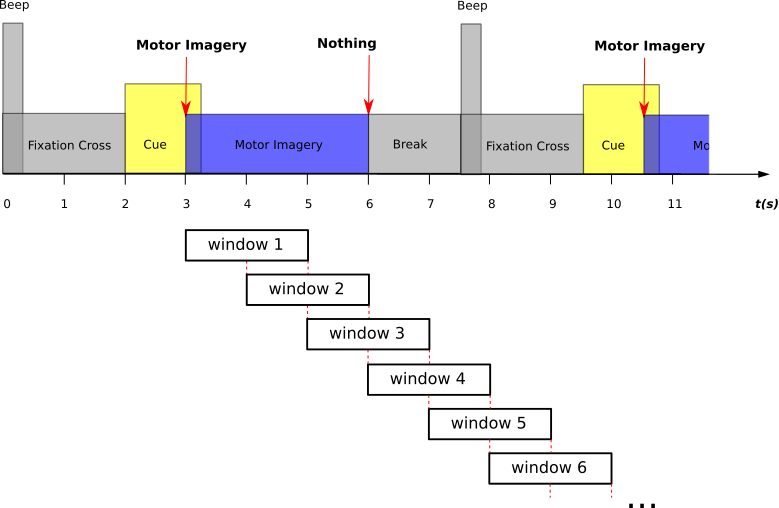}
 \caption{Figure explaining the introduction of the \textit{nothing} event and the sliding windows in the BCI Dataset 2a (BNCI2014001)~\cite{tangermann-muller-etal:12}. In this example, we use a window of 2 seconds with an overlapping of $50\%$}
  \label{fig1}
\end{figure}

To enable the framework to operate in pseudo-online mode, an extension of the methodology is required, taking into account that tasks can vary in duration. This extension is achieved by employing a sliding window approach. In the MOABB framework, a single trial is extracted for each task performed by the subject, resulting in a pure signal consisting of only one epoch extracted from each task. However, in order to achieve a performance evaluation that closely resembles a real BCI application, it is necessary to transform the dataset into an asynchronous format using an overlapping sliding windows approach.
The idea is to select a sliding windows with size $T$ that is smaller than the total time of the task and run it on the continuous data with a step size, which is controlled using the overlapping parameter (see \ref{fig1}). In general, the optimal values for the length of the sliding window and the overlapping is a trade-off between accuracy and response speed. 

The sliding window approach can be seen as a data augmentation procedure. This approach actually generates a significantly greater number of trials per class compared to the original method~\cite{lashgari-liang-etal:20}.

Implementing the sliding windows approach introduces the challenge of generating windows that contain a mixture of events, as the sliding operation spans across the entire continuous data. To deal with this issue, we had to assign a unique label to these events, in order to create a dataset that is compatible with most machine learning algorithm. The assignment of this unique label is based on the percentage of data contained within a specific mixed window. We consider that windows are small enough so that only two events can appear in a window. Let us call $a$ and $b$ respectively the percentages of the window with the first and second events:
\begin{itemize}
    \item If $a>b$, we assign the label of the initial event.
    \item If $a \leq b$, we assign the label of the last event. This label is used in case of equality because the subject intent is to perform the new task.
\end{itemize}

\begin{figure}[!ht]
 \centering
 \includegraphics[width=\linewidth]{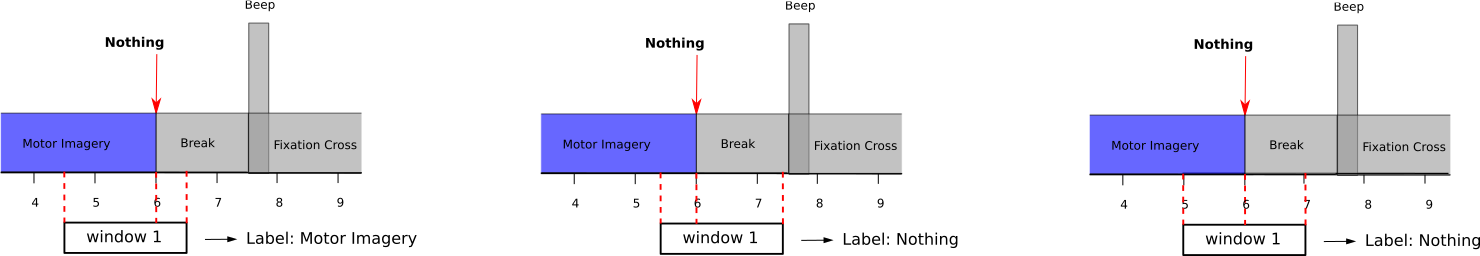}
 \caption{The different ways to treat the windows that contain two events.}
  \label{fig2}
\end{figure}

A detailed representation of this procedure can be found in~\ref{fig2}.

We enforce the evaluation process to include all the different tasks plus the new \textit{nothing} task, generating a non binary classification. This problem will be taken in account in~\ref{metric}.

\subsection{Evaluation}
Having split the continuous data using the sliding window approach, we are now ready to evaluate the performance of several algorithms on these modified datasets. In this section, we discuss the metrics used as well as the possible different evaluations types: Within-Session and Cross-Session.

\subsubsection{Metrics}
\label{metric}
The transformation of the dataset introduces an important issue: the transformed dataset is strongly unbalanced with respect to the \textit{nothing} event. Furthermore, the transformation always introduces a new class, so that we have to deal with non-binary classification throughout our processing.

Different solutions are possible in such situations. Collecting a large amount of data for BCI purposes can be expensive and time-consuming, making data cancellation (randomly delete samples from the majority class to balance the class distribution) an impractical option. To solve this problem, we adopted a metric that is reliable with unbalanced datasets~\cite{thomas-dyson-etal:13}. The standard measure used in BCI is accuracy, which gives reliable results for balanced datasets. When this condition fails, accuracy produces an overly optimistic performance estimation.
While such a metric is perfectly adequate to evaluate the performance in a synchronized BCI where usually the datasets are balanced, it no longer is in an asynchronous setting.
In such a situation, the BCI literature recommends the use of Cohen's Kappa coefficient~\cite{thomas-dyson-etal:13, cohen:60}.
There are however other measures to deal with unbalanced datasets~\cite{zhu:20} such as Matthews Correlation Coefficient (MCC)~\cite{matthews:75}, which was introduced by Matthews in the case of binary classification~\cite{matthews:75} and generalized to multi-class problems~\cite{gorodkin:04}.
Recent research has shown that Cohen's Kappa and MCC performance measures are very similar in most situations, but may differ in others. This leads to anomalous performance for Cohen's Kappa in certain situations, which is why we preferred to use MCC~\cite{delgado-tibau:19, chicco-warrens-etal:21}.
In addition, MCC has been shown to be much more informative than several metrics including ROC-AUC in binary classification~\cite{chicco-jurman:23}.
In the case of binary classification MCC is defined as:

\begin{equation}
    MCC=\frac{TP \times TN-FP \times FN}{\sqrt{(TP+FP)(TP+FN)(TN+FP)(TN+FN)}} \;,
\end{equation}
where $TP$, $TN$, $FP$ and $FN$ are respectively the number of true positives, true negatives, false positives and false negatives defined using the confusion matrix. MCC lies in the range [-1, +1], with -1 and +1 being reached respectively in case of perfect misclassification and perfect classification. MCC = 0 is the expected value for the coin tossing classifier~\cite{chicco-jurman:20}. 
We decided to use the normalized version of MCC in the framework, the nMCC is defined as $\mathrm{nMCC}=\frac{\mathrm{MCC}+1}{2}$.

Such normalization projects the original range [-1,+1] into the interval [0, +1], where +1 correspond to a perfect classification while $nMCC = 0.5$ is for prediction similar to random guessing. Identical considerations apply to its extension to multi label classification. 

The performance of a BCI system can also be evaluated by how much information can be transferred without committing errors in a specific time frame, i.e., the bit-rate of the system.
Usually, the information transfer rate (ITR) is used, of which two definitions have been formulated: the first, which was proposed by Wolpaw \textit{et al}~\cite{wolpaw-ramoser-etal:98} includes the use of accuracy in its definition and thus is based on the same assumptions as accuracy. In our situation, we therefore decided to adopt the definition of ITR given by Nykopp~\cite{nykopp-etal:01, sadeghi-maleki:19}, that is based on the concept of Mutual Information (MI)~\cite{kraskov-stogbauer-etal:04}.
The MI between two discrete random variables $X$ and $Y$ is defined as 

\begin{equation}
    MI(X ; Y)=\sum_{y \varepsilon Y} \sum_{x \varepsilon X} p(x, y) \log \frac{p(x, y)}{p(x) p(y)}
\end{equation}
where $p(x, y)$ is the joint probability of realizing events $x$ and $y$ simultaneously and $p(x)$ and $p(y)$ is the probability associated with the individual variables. The logarithm used in this context is base two, as information is measured and conveyed in bits. ITR is then defined as the amount of information transmitted per minute (bits/minute)

\begin{equation}
    ITR = MI(X, Y) \frac{60}{T}
\end{equation}
where $T$ (seconds/symbol) is the time in seconds needed to transmit a symbol, in our case to select a task.
Similar considerations apply to its extension to multi label classification.

\subsubsection{Within-Session Evaluation Procedure}
The Within-Session evaluation procedure involves evaluating performance directly within the same session of a certain subject. The current evaluation method employed in MOABB utilizes a 5-fold Cross Validation. However, in the Pseudo-Online extension, we decided to not use Cross Validation as our objective is to explicitly preserve the causal relationship within the data. Therefore, we enforced that the test dataset temporally follows the training portion. For the same reason, we also decided to not shuffle the data. 
Since, in the datasets considered, there is not a predefined split in the training and test part, we followed the state-of-the-art procedure, which is to split the dataset into a training dataset containing the first $80\%$ windows and a test dataset containing the $20\%$ remaining ones. For the hyper-parameter, we used a 5-fold cross validation on the training dataset.

\subsubsection{Cross-Session Evaluation Procedure}
The Cross-Session evaluation procedure focuses on a single subject and incorporates all sessions except one for the training phase, while utilizing the remaining session for the testing phase. This approach is carried out with a Leave One Out Cross Validation.

In order to allow a complete fairness of the approach, we performed a Nested Cross Validation when the hyper parameter tuning was necessary~\cite{cawley-talbot:10}. However, it is noticeable that the performance obtained using the nested approach is statistically similar to that obtained with the less computationally intensive flat cross-validation approach, a finding that is aligned with some recent research~\cite{wainer-cawley:21}. We chose to produce the Nested Cross Validation results, but those obtained using the flat cross-validation are given for comparison in the appendix in Table~\ref{table:Cross_Flat}.

\subsubsection{Pipelines Considered}
We considered different state of the art pipelines for Motor Imagery classification in BCI described in~\ref{table:Pipeline}.  This list covers the algorithms that have shown good classification performance in the last 15 years. This list is not intended as definitive but rather as a starting point: each research group will be able to add its algorithm to the dashboard, after testing it/them in the same setting. 

To perform this comparison, it was also necessary to extend MOABB, currently based on the sole scikit-learn library to enable it to use the Deep Learning (DL) frameworks of TensorFlow and Keras.

In recent years, DL algorithms became increasingly popular for solving extremely complex problems that could not be solved by traditional Machine Learning (ML) approaches.  The popularity of such algorithms is due to recent successes in a wide variety of fields, from Natural Language processing~\cite{collobert-weston-etal:11} to image recognition~\cite{krizhevsky-sutskever-etal:17}. Only recently have such algorithms begun to be applied for BCI classification. 

Conventional ML -- non DL -- algorithms are not suitable to process directly the raw data. There is usually a feature extraction step that is designed with some domain expertise. On the contrary, DL models have shown remarkable capabilities in automatically learning and extracting relevant features from raw data, such as EEG signals. In addition, these models proved to be particularly adaptable and generelizable to new subjects and sessions. 
Despite their potential, DL models are not extensively used in the field of BCI due to several challenges: they typically demand a substantial volume of training data, which can be difficult and costly to acquire in BCI due to the specialized equipment and expertise required. Additionally, they are often perceived as black boxes, lacking interpretability and making it challenging to understand the decision-making process. In BCI applications, interpretability is essential for fostering user trust, gaining clinical acceptance, and enabling effective feedback mechanisms.

To replicate the state of the art, we used the Keras~\cite{chollet-etal:15} framework and the KerasClassifier function from the SciKeras package. With this library, it is possible to create a deep learning architecture and convert it into a scikit-learn pipeline that can be integrated directly into the standard MOABB framework.
In order to make the results more stable, every deep learning pipeline is preceded by a standardization step which puts each channel to zero mean and unit standard deviation. We also apply a re-sampling procedure to ensures that each architecture incorporates a temporal filter that is aligned with the implementation provided in the state-of-the-art techniques. 
Moreover, the same sliding window parameters were used for all algorithms in order to allow for a fair comparison. Details on the DL hyper parameters are given in~Table~\ref{table:pipeline_parameter_DL}.

\begin{table*}[!ht]
\begin{center}
\caption{Pipelines considered in this study.} 
\label{table:Pipeline} 
\footnotesize
\begin{tabularx}{\linewidth}[t]{>{\hsize=.7\hsize}cLL>{\hsize=.35\hsize}c}
\hline
\textbf{Name Pipeline} & \textbf{Feature Extraction} & \textbf{Classifier} & \textbf{References} \\
\hline 
MDM & Spatial Covariance estimated with Sample Estimator & Mean Distance to Mean (MDM) & \cite{barachant-bonnet-etal:10} \\
Cov + EN & Spatial Covariance estimated with Sample Estimator mapped to Tangent Space & Optimized Elastic Network (EL) & \cite{corsi-chevallier-etal:22} \\
FgMDM & Spatial Covariance estimated with Sample Estimator & Minimum Distance to Mean with geodesic filtering (FgMDM) & \cite{barachant-bonnet-etal:10} \\
TANG + SVM & Spatial Covariance estimated with Sample Estimator mapped to Tangent Space & Optimized Support Vector Machine (SVM) & \cite{barachant-bonnet-etal:10} \\
AUG + TANG + SVM & Augmented Spatial Covariance estimated with Sample Estimator mapped to Tangent Space & Optimized SVM & \cite{carrara-papadopoulo:23} \\
CSP + LDA & Common Spatial Patterns (CSP) & Optimized shrinkage Linear Discriminant Analysis (LDA) & \cite{lotte-bougrain-etal:18} \\
CSP + RF & CSP & Optimized Random Forest (RF) & \cite{lotte-bougrain-etal:18} \\
CSP + SVM & CSP & Optimized SVM & \cite{lotte-bougrain-etal:18} \\
AR + SVM & Autoregressive Coefficient & Optimized SVM & \cite{al-faiz-al-hamadani:19} \\
AR + LR & Autoregressive Coefficient & Optimized Linear Regression (LR) & \cite{al-faiz-al-hamadani:19} \\
FBCSP+LDA & Filter Bank Common Spatial Patterns (FBCSP) & Optimized Shrinkage LDA & \cite{ang-chin-etal:08, lotte-bougrain-etal:18} \\
FBCSP+SVM & FBCSP & Optimized SVM & \cite{ang-chin-etal:08} \\
FBCSP+MLP & FBCSP & MLP & \cite{ang-chin-etal:08} \\
FBCSP+RF & FBCSP & Optimized RF & \cite{ang-chin-etal:08} \\
ShallowConvNet & Standardized and resample EEG signal at 250Hz & CNN & \cite{schirrmeister-springenberg-etal:17} \\
DeepConvNet & Standardized and resample EEG signal at 250Hz & CNN & \cite{schirrmeister-springenberg-etal:17} \\
EEGNet 8 2 & Standardized and resample EEG signal at 128Hz & CNN with architecture EEGNet & \cite{lawhern-solon-etal:18} \\
EEGTCNet & Standardized and resample EEG signal at 250Hz & CNN with architecture EEGTCNet & \cite{ingolfsson-hersche-etal:20} \\
EEGITNet & Standardized and resample EEG signal at 128Hz & CNN with architecture EEGITNet & \cite{salami-andreu-perez-etal:22} \\
EEGNeX 8 32 & Standardized and resample EEG signal at 128Hz & CNN with architecture EEGNeX & \cite{chen-teng-etal:22} \\

\end{tabularx}
\end{center}
\end{table*}

\begin{table}[!ht]
\begin{center}
    \caption{Common parameters for DL pipelines.}
\label{table:pipeline_parameter_DL}
\resizebox{0.5\linewidth}{!}{\begin{tabular}{c|c}
  \hline
  \textbf{Parameter}  &  \textbf{Value} \\ \hline
    Epoch & 300 \\ \hline
    Batch Size & 64\\ \hline
    Validation Split & 0.2 \\ \hline
    Loss & Sparse Categorical Crossentropy \\ \hline
    Optimizer & Adam \\
    & Learning Rate = 0.0009 \\ \hline
    Callbacks ES & Early Stopping \\
    & Patience = 75 \\
    & Monitor = Validation Loss \\ \hline
    Callbacks LR & ReduceLROnPlateau \\
    & Patience = 75 \\
    & Monitor = Validation Loss \\
    & Factor = 0.5 \\ \hline
\end{tabular}
}
\end{center}
\end{table}
When possible, the hyper-parameters of the classification models were optimized using a Grid Search procedure. We did not create an ablation study for the DL models since we faithfully reproduced the architectures proposed in the respective references.

\section{Results}
\label{Results}
In this section, we report the performance results obtained with the pipelines considered. Ultimately, to validate the robustness and validity of our \textit{pseudo-online} approach, we tested the algorithms on different datasets, subjects and tasks.

\subsection{Paradigm}
The sliding window is defined to have a 2\,s duration windows with a $50\%$ overlapping.

Except for the filter-bank base algorithms, we applied on all datasets a standard preprocessing -- for the motor imagery task -- band-pass filter in the region [8; 30] Hz. 
For pipelines based on the Filter Bank paradigm, we used 6 different non overlapping windows in order to filter the EEG signal into 8--12 Hz, 12--16 Hz, 16--20 Hz, 20--24 Hz, 24--28 Hz, 28--35 Hz.

\subsection{Pseudo Online Evaluation}

\subsubsection{Within-Session Evaluation}
% Report table with results and 3 figure for every dataset
% Meta Analysis Plot
We give the results of the \textit{pseudo-online} evaluation using the within-session methodology in Table~\ref{table:Within}.
These results are also displayed in the appendix (\ref{fig:BNCI2014001_WS}, \ref{fig:BNCI2014002_WS}, \ref{fig:BNCI2014004_WS} and \ref{fig:BNCI2015001_WS}), showing also a detailed study of the statistical significance.

\begin{table*}[!ht]
\caption{Performance Pseudo Online Within-Session Evaluation. Results for the DL architecture are listed after the two line.}
\label{table:Within}
\centering
\resizebox{0.8\linewidth}{!}{\begin{tabular}{c|c|c|c|c}
Pipeline &                BNCI2014002       & BNCI2014004           & BNCI2015001       & BNCI2014001\\
\hline MDM &              $0.66 \pm 0.09$ & 0.63 $\pm$ 0.06 &      0.72 $\pm$ 0.07 & $0.67 \pm 0.05$  \\
\hline Cov + EN &         $0.65 \pm 0.09$ & 0.62 $\pm$ 0.08 &      0.73 $\pm$ 0.08 & $0.69 \pm 0.07$  \\
\hline FgMDM &            $0.67 \pm 0.09$ & 0.64 $\pm$ 0.06 &      0.74 $\pm$ 0.07 & $0.70 \pm 0.07$  \\
\hline TANG + SVM &       $0.69 \pm 0.10$ & 0.61 $\pm$ 0.08 &      0.74 $\pm$ 0.08 & $0.70 \pm 0.08$  \\
\hline AUG + TANG + SVM & $\textbf{0.70} \pm \textbf{0.10}$ & \textbf{0.66} $\pm$ \textbf{0.09} &      \textbf{0.76} $\pm$ \textbf{0.08} & $0.70 \pm 0.07$  \\
\hline CSP + LDA &        $0.58 \pm 0.10$ & 0.60 $\pm$ 0.08 &      0.65 $\pm$ 0.09 & $0.57 \pm 0.05$  \\
\hline CSP + RF &         $0.59 \pm 0.10$ & 0.61 $\pm$ 0.06 &      0.65 $\pm$ 0.08 & $0.56 \pm 0.05$  \\
\hline CSP + SVM &        $0.59 \pm 0.10$ & 0.60 $\pm$ 0.08 &      0.65 $\pm$ 0.08 & $0.57 \pm 0.05$  \\
\hline FBCSP+LDA &        $0.69 \pm 0.10$ & 0.65 $\pm$ 0.09 &      0.74 $\pm$ 0.08 & $\textbf{0.71} \pm \textbf{0.06}$  \\
\hline FBCSP+SVM &        $0.69 \pm 0.11$ & 0.64 $\pm$ 0.09 &      0.74 $\pm$ 0.09 & $0.70 \pm 0.07$  \\
\hline FBCSP+MLP &        $0.67 \pm 0.11$ & 0.65 $\pm$ 0.08 &      0.72 $\pm$ 0.09 & $0.69 \pm 0.06$  \\
\hline FBCSP+RF &         $0.69 \pm 0.10$ & 0.63 $\pm$ 0.08 &      0.73 $\pm$ 0.09 & $0.68 \pm 0.07$  \\ \hline
\hline ShallowConvNet &   $0.64 \pm 0.10$ & 0.58 $\pm$ 0.06 &      0.73 $\pm$ 0.08 & $0.69 \pm 0.08$  \\
\hline DeepConvNet &      $0.61 \pm 0.09$ & 0.58 $\pm$ 0.05 &      0.63 $\pm$ 0.08 & $0.61 \pm 0.07$  \\
\hline EEGNet 8 2 &       $0.65 \pm 0.08$ & 0.59 $\pm$ 0.07 &      0.73 $\pm$ 0.07 & $0.67 \pm 0.07$  \\
\hline EEG ITNet &        $0.61 \pm 0.08$ & 0.57 $\pm$ 0.06 &      0.68 $\pm$ 0.08 & $0.64 \pm 0.08$  \\
\hline EEG TCNet &        $0.64 \pm 0.10$ & 0.58 $\pm$ 0.08 &      0.70 $\pm$ 0.08 & $0.66 \pm 0.07$  \\
\hline EEGNeX 8 32 &      $0.58 \pm 0.08$ & 0.56 $\pm$ 0.06 &      0.66 $\pm$ 0.07 & $0.61 \pm 0.07$  \\
\hline 
\end{tabular}
}
\end{table*}

\subsubsection{Cross-Session Evaluation}
We give the results of the \textit{pseudo-online} evaluation using the Cross-Session methodology in Table~\ref{table:Cross_Nested}. 
These results are also displayed in the appendix (\ref{fig:BNCI2014001_CS}, \ref{fig:BNCI2014004_CS} and \ref{fig:BNCI2015001_CS}), showing also a detailed study of the statistical significance.
%\TPcommentWhy{ is there on benchmark less than for the within session case?}

\begin{table*}[!ht]
\caption{Performance Pseudo Online Cross-Session Evaluation Using Nested Cross Validation. Results for the DL architecture are listed after the two line.}
\label{table:Cross_Nested}
\centering
\resizebox{0.65\linewidth}{!}{\begin{tabular}{c|c|c|c}
Pipeline &                BNCI2014004           & BNCI2015001       & BNCI2014001\\
\hline MDM &              0.61 $\pm$ 0.06 &      0.67 $\pm$ 0.07 & $0.65 \pm 0.05$  \\
\hline Cov + EN &         0.59 $\pm$ 0.07 &      0.69 $\pm$ 0.07 & $0.67 \pm 0.06$  \\
\hline FgMDM &            0.62 $\pm$ 0.06 &      0.69 $\pm$ 0.06 & $0.67 \pm 0.06$  \\
\hline TANG + SVM &       0.57 $\pm$ 0.07 &      0.68 $\pm$ 0.07 & $0.65 \pm 0.08$  \\
\hline AUG + TANG + SVM & \textbf{0.63} $\pm$ \textbf{0.08} &      \textbf{0.72} $\pm$ \textbf{0.06} & $\textbf{0.69} \pm \textbf{0.06}$  \\
\hline CSP + LDA &        0.58 $\pm$ 0.07 &      0.61 $\pm$ 0.07 & $0.57 \pm 0.05$  \\
\hline CSP + RF &         0.58 $\pm$ 0.05 &      0.61 $\pm$ 0.07 & $0.56 \pm 0.04$  \\
\hline CSP + SVM &        0.57 $\pm$ 0.07 &      0.61 $\pm$ 0.07 & $0.56 \pm 0.05$  \\
\hline FBCSP+LDA &        0.62 $\pm$ 0.07 &      0.70 $\pm$ 0.07 & $0.68 \pm 0.06$  \\
\hline FBCSP+SVM &        0.61 $\pm$ 0.07 &      0.68 $\pm$ 0.08 & $0.67 \pm 0.06$  \\
\hline FBCSP+MLP &        0.62 $\pm$ 0.06 &      0.68 $\pm$ 0.08 & $0.65 \pm 0.06$  \\
\hline FBCSP+RF &         0.61 $\pm$ 0.07 &      0.67 $\pm$ 0.08 & $0.66 \pm 0.06$  \\ \hline
\hline ShallowConvNet &   0.55 $\pm$ 0.05 &      0.70 $\pm$ 0.07 & $0.69 \pm 0.06$  \\
\hline DeepConvNet &      0.56 $\pm$ 0.04 &      0.64 $\pm$ 0.05 & $0.62 \pm 0.07$  \\
\hline EEGNet 8 2 &       0.56 $\pm$ 0.05 &      0.70 $\pm$ 0.06 & $0.67 \pm 0.07$  \\
\hline EEG ITNet &        0.55 $\pm$ 0.05 &      0.68 $\pm$ 0.07 & $0.65 \pm 0.07$  \\
\hline EEG TCNet &        0.56 $\pm$ 0.06 &      0.70 $\pm$ 0.06 & $0.66 \pm 0.06$  \\
\hline EEGNeX 8 32 &      0.55 $\pm$ 0.05 &      0.65 $\pm$ 0.06 & $0.61 \pm 0.06$  \\
\hline 
\end{tabular}
}
\end{table*}

\section{Discussion}
\label{Discussion}
The numerous results performed on different datasets with different classification tasks, both binary and multi-class, showed a good alignment with state-of-the-art results. Overall, the method that shows the best performance is the augmented covariance method with classification using SVM on the tangent space~\cite{carrara-papadopoulo:23}. In some situations, the FBCSP-based algorithms also obtain comparable results $($BNCI2014001, BNCI2014002$)$, while in the remaining considered datasets $($BNCI2014004, BNCI2015001$)$, the performance produced by the augmented covariance method appears to be superior.

However, the augmented covariance method depends on the selection of two hyper parameters with a grid search which is computationally intensive due to the increasing size of the augmented covariance matrix with the order parameter~\cite{you-park:22}. 
%However, this issue can be overcome in a possible online application, in fact it is enough to ask the subject to perform a calibrating session and only after training the hyper parameters use the online algorithm. 

In this analysis, it is shown that in general DL algorithms have lower performance than standard machine learning algorithms. One possible explanation is that more data are needed to train the models efficiently, eventually using data augmentation procedures, such as introducing random Gaussian noise on the data or time inversion procedures~\cite{rommel-paillard-etal:22}.

\subsection{Comparison Pseudo Online vs Offline Evaluation}
In general, it is observed that the performance achieved through pseudo-online evaluation is lower compared to offline evaluation. This can be attributed to several factors. Firstly, in order to enable real-time applications, the duration of epoch sliding window is typically reduced, which can impact the accuracy of classification. Secondly, the introduction of the \textit{nothing} event introduces an additional class to the already complex classification task. The \textit{nothing} event encompasses a wide range of mental phenomena, resulting in high variability within this class, which in turn affects the overall performance of the classification task.
%\TP{}{10 to 15\% (I put numbers from memory}, put more exact ones)

\subsubsection{Within-Session Evaluation}
% Report table with results and 3 figure for every dataset
% Meta Analysis Plot
We want to emphasize the change in performance and ranking for the best algorithms using the standard offline against the pseudo-online using the Within-Session evaluation methodology, both using nMCC as metric. The performance of the \textit{offline} methodology turns out to be much better compared to the \textit{pseudo-online} evaluation as shown in Tables~\ref{table:Within_Offline} and~\ref{table:Within}.

\begin{figure*}[!ht]  
    \centering
    \centering
     \subfloat[]{%
            \includegraphics[width=0.45\linewidth]{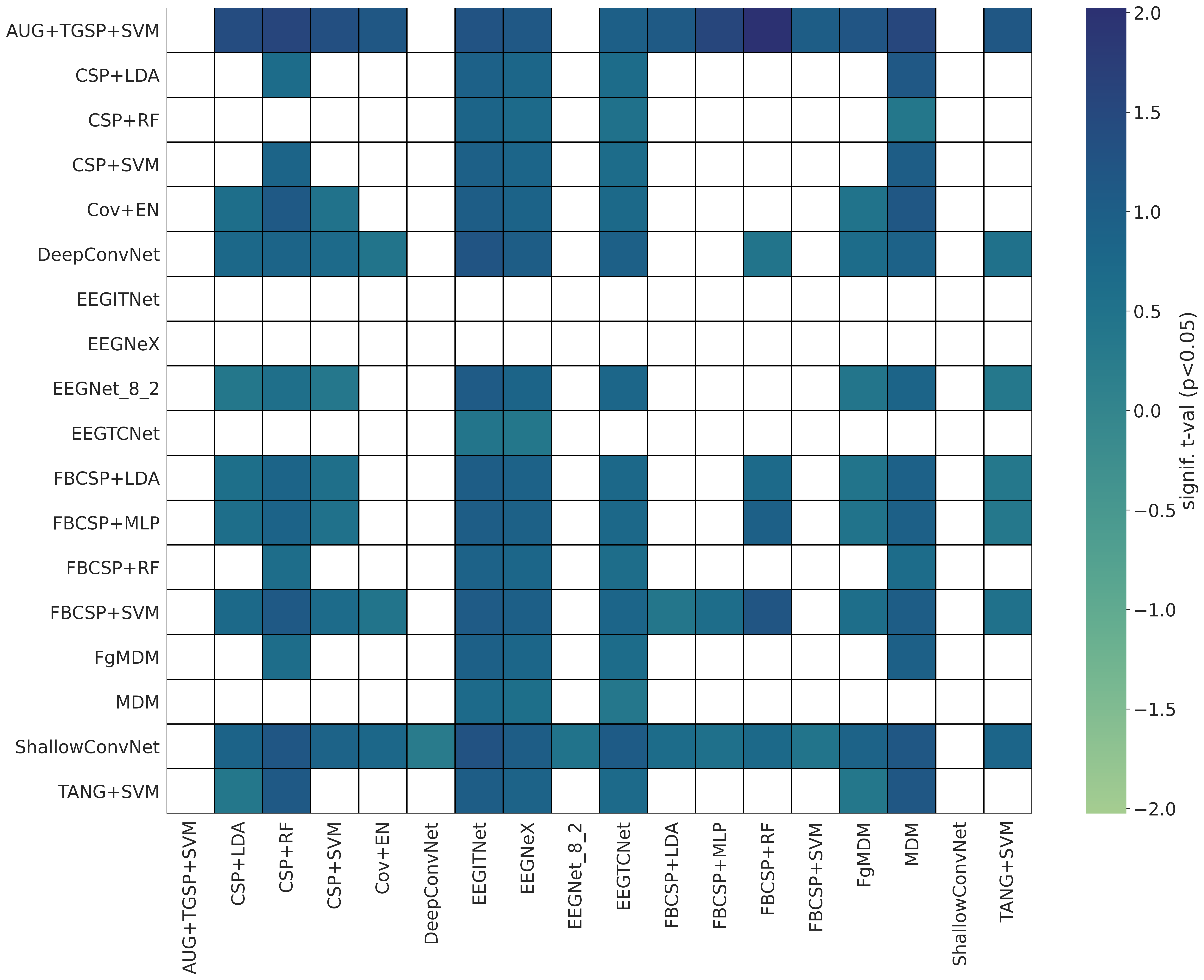}}
             \hfill
     \subfloat[]{%
            \includegraphics[width=0.45\linewidth]{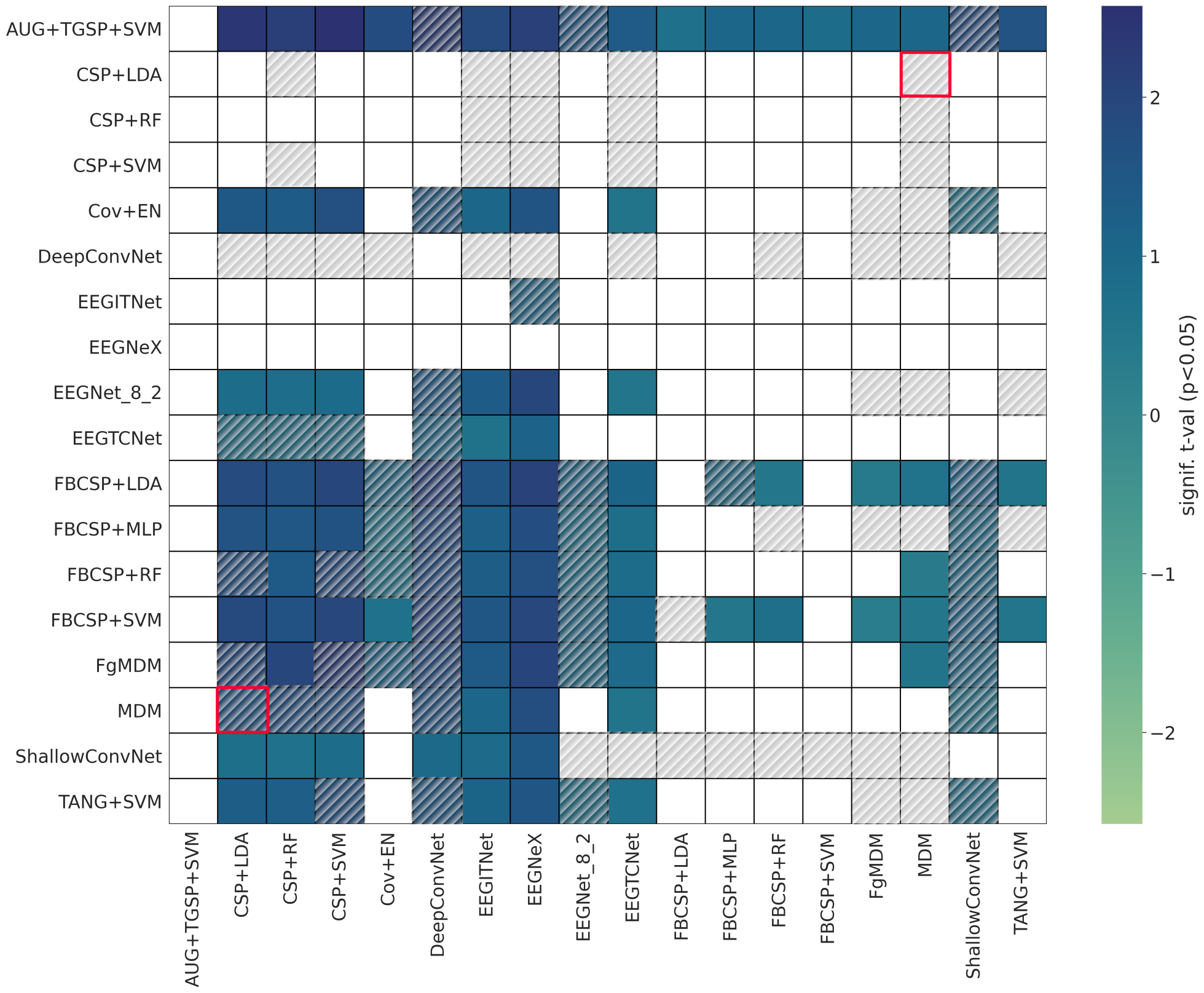}} 
    \caption{Result for 3 task classification using results of dataset BNCI2015001, BNCI2014002 and BNCI2014004 using Within-Session evaluation. Plot (a) shows the meta analysis of the different methods considered using offline evaluation. Plot (b) shows the meta analysis of the different methods considered using pseudo online evaluation.
    The grey zone is where we find a statistical significant difference between the two evaluations. Red boxes indicate the behaviour analyzed in~\ref{fig:Meta_comparison_WS}, which plots the significance that the algorithm on the y-axis is better than the one on the x-axis. The color represents the significance level of the difference of accuracy, in terms of t-values, and we show only the significant interactions ($p < 0.05$).
    %\TPcomment{I corrected the (A) and (B) into (a) and (b) but the right way is to use subcaption and labels. Same thing in other Figures 5, 7 8 9 10 11 12 13 14, in which I did not made the corrections.}
    }
    \label{fig:Comparison_WS}
\end{figure*}

\begin{figure}[!ht]  
    \centering
    \centering
     \subfloat[]{%
            \includegraphics[width=0.45\linewidth]{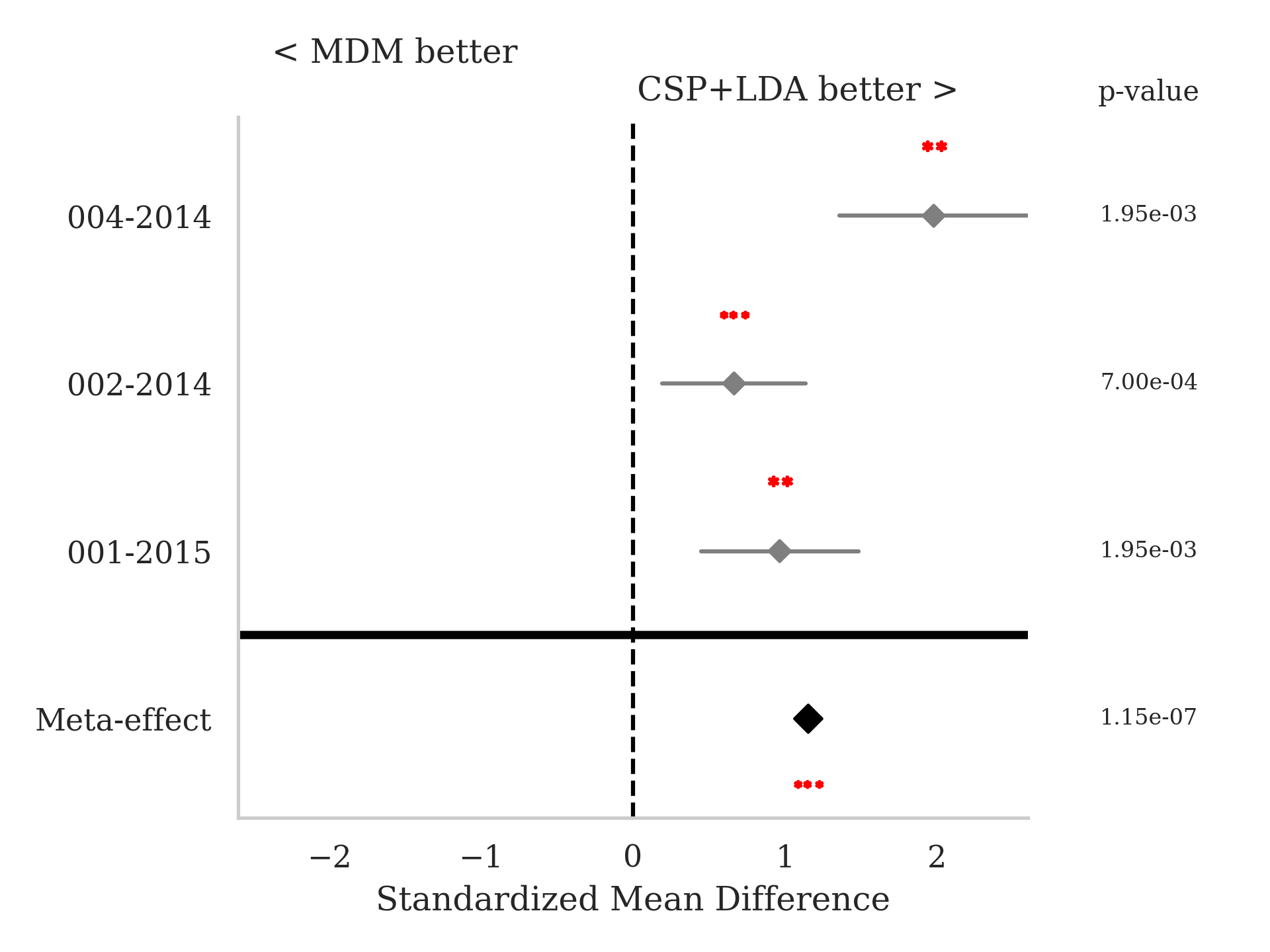}}
             \hfill
     \subfloat[]{%
            \includegraphics[width=0.45\linewidth]{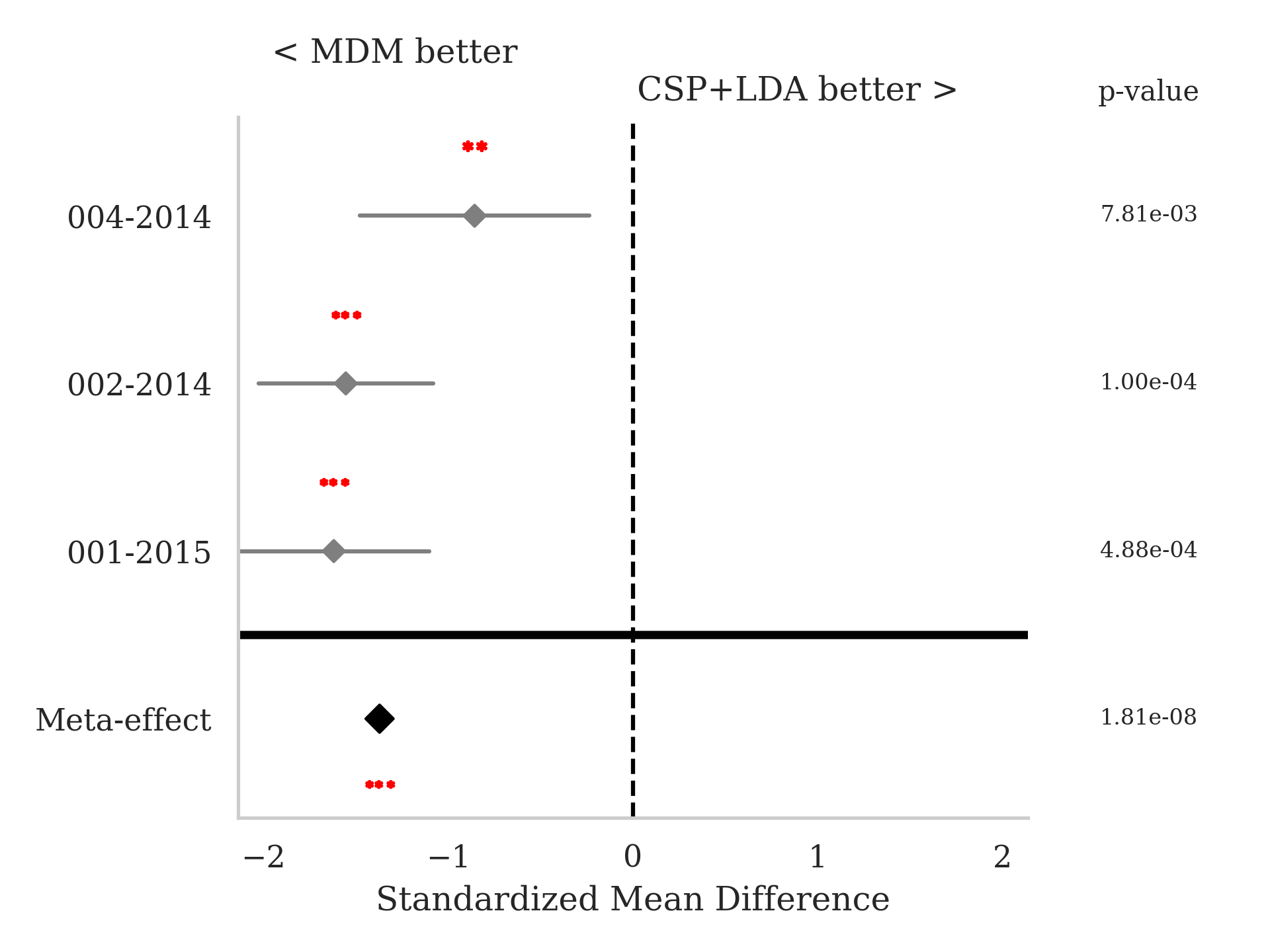}} 
    \caption{Result for Meta Analysis of CSP+LDA vs MDM on 3 task classification using results of dataset BNCI2015001, BNCI2014002 and BNCI2014004 using Within-Session evaluation. Plot (A) shows the meta analysis using offline evaluation. Plot (B) shows the meta analysis using pseudo online evaluation. We show the standardized mean differences, while p-values are computed as one-tailed Wilcoxon signed-rank test for the hypothesis given as title of the plot and the gray bar  denote $95\%$ interval. Here, * stands for $p < 0.05$, ** for $p < 0.01$, and *** for $p < 0.001$.
    }
    \label{fig:Meta_comparison_WS}
\end{figure}

A detailed analysis of Figure~\ref{fig:Comparison_WS} reveals several noteworthy differences indicated by the gray regions in Figure~\ref{fig:Comparison_WS}(b). Firstly, the ACM methodology outperforms the state of the art approaches in both \textit{offline} and \textit{pseudo online} evaluations. Secondly, some DL algorithms seem to be more stable in the \textit{pseudo online} approach for the BNCI2014001 dataset. This can be attributed to the utilization of sliding windows as a data augmentation technique, highlighting the significant dependence of -- at least some -- DL algorithms on large amounts of data. The ranking of certain algorithms is completely opposite: the red box in Figure~\ref{fig:Comparison_WS}(b) representing the pipelines "CSP+LDA" and "MDM" are an example of such a ranking change. Figure~\ref{fig:Meta_comparison_WS} presents a meta-analysis of these two pipelines. The results demonstrate the superiority of the "CSP+LDA" pipeline in the \textit{offline} evaluation that is completely reversed in the \textit{pseudo-online} approach.

\subsubsection{Cross-Session Evaluation}
We made the same analysis of change in performance and ranking for the best algorithms for the Cross-Session evaluation. To compare the performance between \textit{offline} and \textit{pseudo-online} refer to Tables~\ref{table:Cross_Offline} and~\ref{table:Cross_Nested}.

\begin{figure*}[!ht]  
    \centering
    \centering
     \subfloat[]{%
            \includegraphics[width=0.45\linewidth]{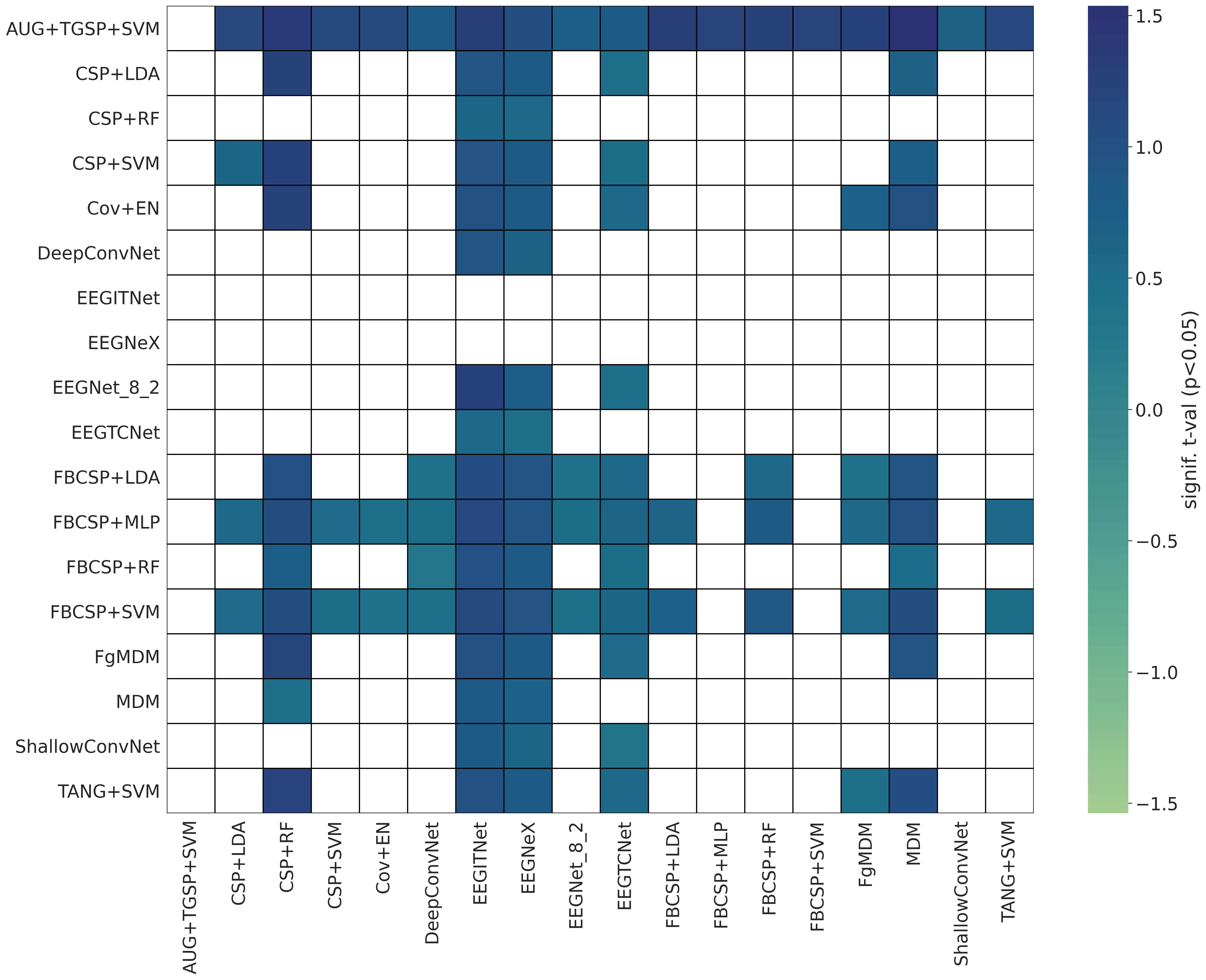}}
             \hfill
     \subfloat[]{%
            \includegraphics[width=0.45\linewidth]{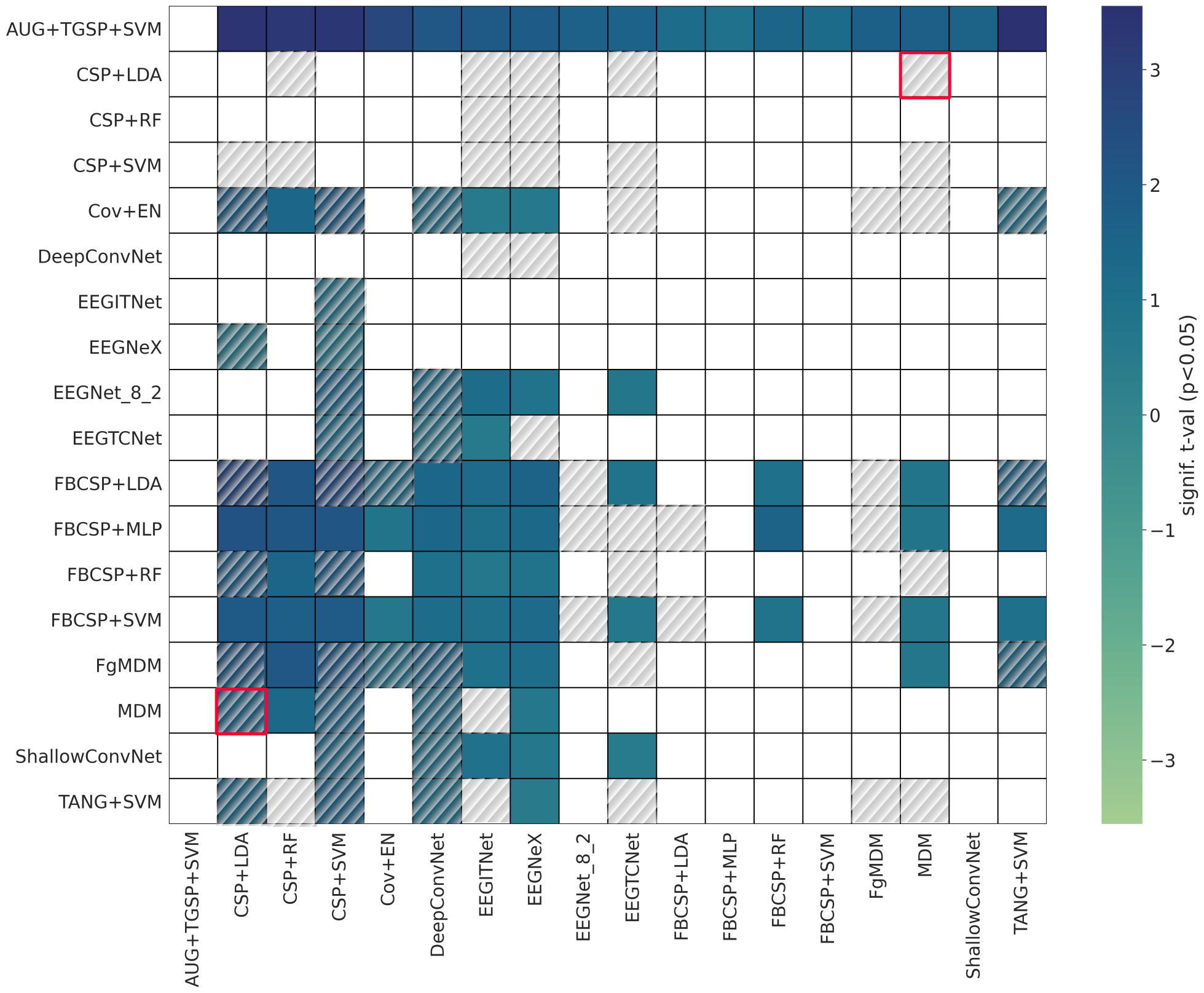}} 
    \caption{Result for 3 task classification using results of dataset BNCI2015001 and BNCI2014004 using Cross-Session evaluation. Plot (a) shows the meta analysis of the different methods considered using offline evaluation. Plot (b) shows the meta analysis of the different methods considered using pseudo online evaluation.
    The grey zone is where we find a statistical significant difference between the two evaluations. Red boxes indicate the behaviour analyzed in~\ref{fig:Meta_comparison_CS}, which plots the significance that the algorithm on the y-axis is better than the one on the x-axis. The color represents the significance level of the difference of accuracy, in terms of t-values, and we show only the significant interactions ($p < 0.05$).
    %\TPcomment{I corrected the (A) and (B) into (a) and (b) but the right way is to use subcaption and labels.}
    }
    \label{fig:Comparison_CS}
\end{figure*}

\begin{figure}[!ht]  
    \centering
    \centering
     \subfloat[]{%
            \includegraphics[width=0.45\linewidth]{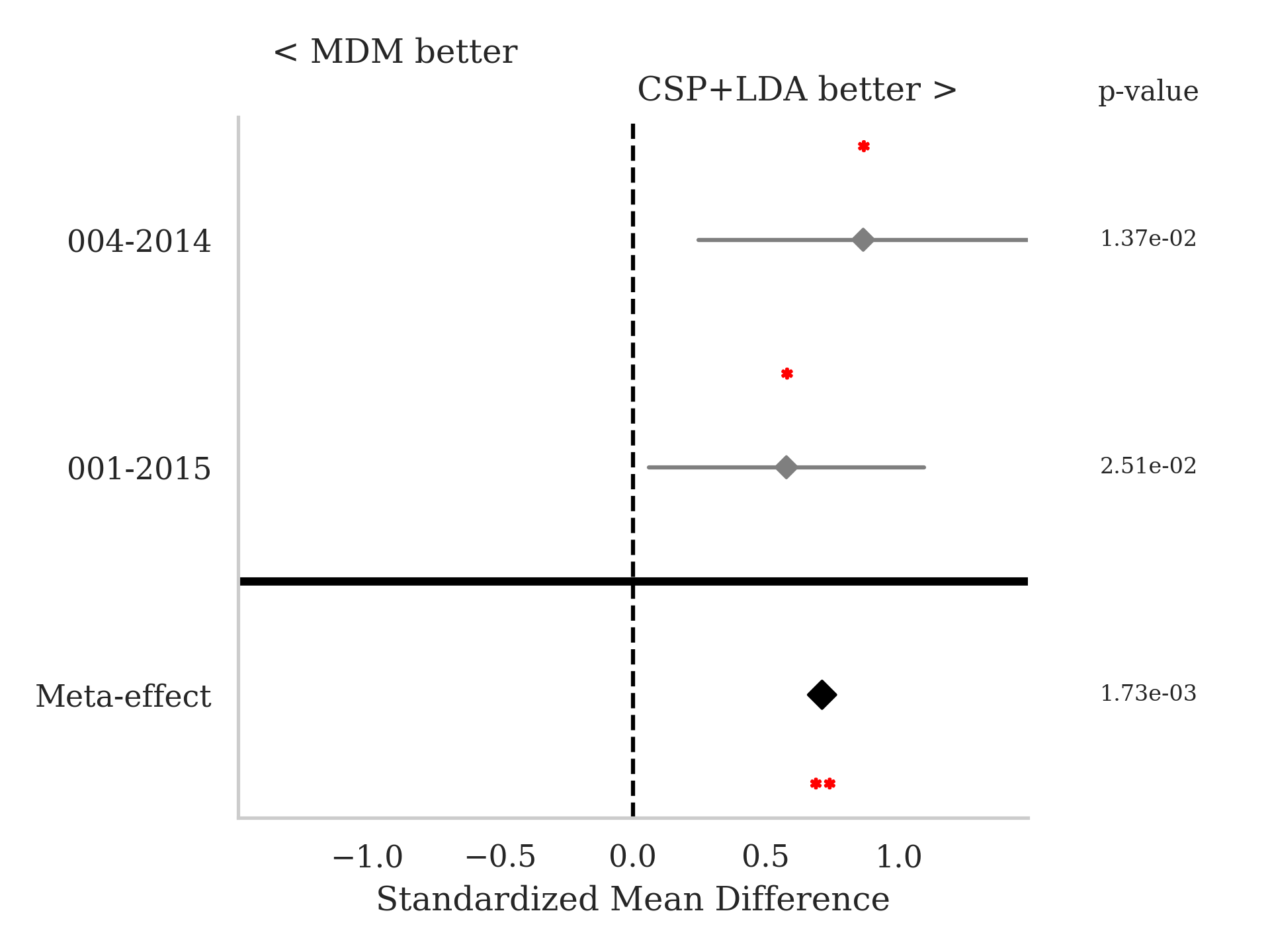}}
             \hfill
     \subfloat[]{%
            \includegraphics[width=0.45\linewidth]{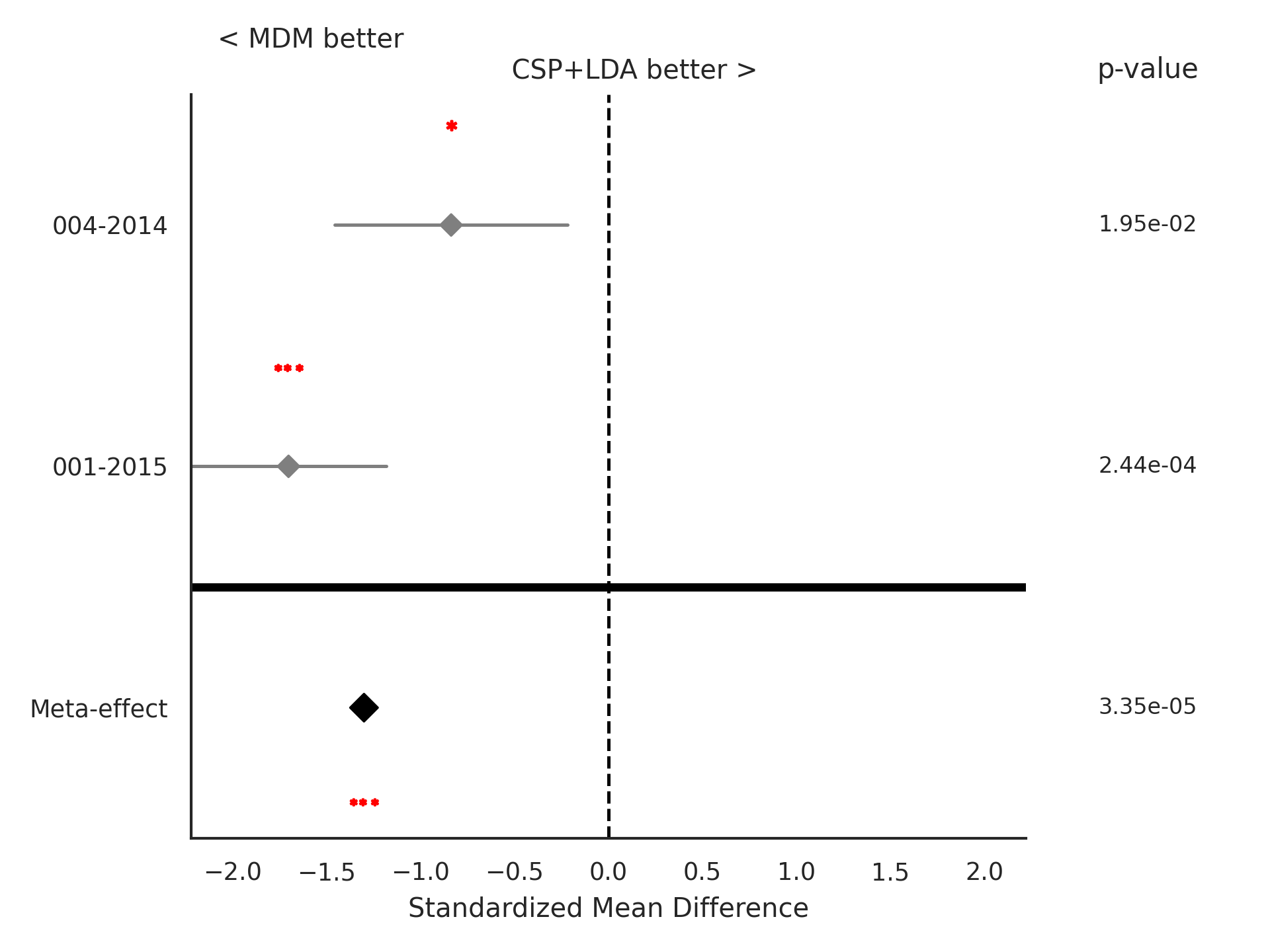}} 
    \caption{Result for Meta Analysis of CSP+LDA vs MDM on 3 task classification using results of dataset BNCI2015001 and BNCI2014004 using Cross-Session evaluation. Plot (a) shows the meta analysis using offline evaluation. Plot (b) shows the meta analysis using pseudo online evaluation. We show the standardized mean differences, while p-values are computed as one-tailed Wilcoxon signed-rank test for the hypothesis given as title of the plot and the gray bar  denote $95\%$ interval. Here, * stands for $p < 0.05$, ** for $p < 0.01$, and *** for $p < 0.001$.
    }
    \label{fig:Meta_comparison_CS}
\end{figure}

A detailed analysis of Figure~\ref{fig:Comparison_CS} reveals several noteworthy differences indicated by the gray regions in Figure~\ref{fig:Comparison_CS}(b). The findings in the Cross-Session case align with our previous observations. In this case also, we noticed a complete reversal in the ranking of certain algorithms compared to the results highlighted in Figure~\ref{fig:Comparison_CS}(b). Specifically, the red boxes representing the pipelines "CSP+LDA" and "MDM" exemplify this ranking discrepancy. To further analyze these pipelines, we conducted a meta-analysis, in Figure~\ref{fig:Meta_comparison_CS}. The outcome of this analysis clearly demonstrates the superiority of the "CSP+LDA" pipeline in the \textit{offline} evaluation, whereas the \textit{pseudo-online} approach completely reverses this superiority.

\section{Conclusion}
\label{Conclusion}
In this research, we introduced an extension of the current MOABB framework in order to provide a framework to test different algorithms in a \textit{pseudo-online} evaluation. In particular, this modification is based on the use of an overlapping sliding windows approach and on the introduction of an \textit{idle} state in the normal Motor Imagery datasets. In order to verify the functioning of such a framework, we tested some of the most efficient algorithms produced in the state-of-the-art of the last 15 years using both ML and DL algorithms. With such a statistical analysis, we show how the augmented covariance approach produces superior performance compared to the state of the art, considering different classification task and different evaluation procedures. We also showed that the efficiency and ranking of the algorithms is highly dependent on the type of analysis -- \textit{offline} or \textit{pseudo-online} -- performed. The \textit{pseudo-online} mode also exhibited some more stable performance for some combinations of DL algorithms and datasets. In conclusion, the ability to analyze the performance of various algorithms in both \textit{offline} and \textit{pseudo-online} modes can significantly accelerate the progress of classification algorithms in the BCI community. By conducting evaluations in \textit{offline} mode initially and then validating the results in \textit{pseudo-online} mode, researchers can effectively enhance the performance of these algorithms. This iterative approach enables the identification of strengths, weaknesses, and areas for improvement, leading to advancements in BCI classification algorithms at a faster pace.

\section*{Acknowledgment}
This work has been partially funded by a EUR DS4H/Neuromod fellowship. The authors are grateful to the OPAL infrastructure from Université Côte d'Azur for providing resources and support. We would like to thanks S. Chevallier for supporting the idea of integrating this approach in MOABB.

\section*{Data and Code Availability}
The code will be soon integrated\footnote{Will be changed to "is integrated" for final publication.} in the MOABB library.

\printbibliography

\section*{Appendix}
%lhrh_within_MDM
\begin{figure*}[!ht]  
    \centering
    \centering
     \subfloat[]{%
            \includegraphics[width=0.45\linewidth]{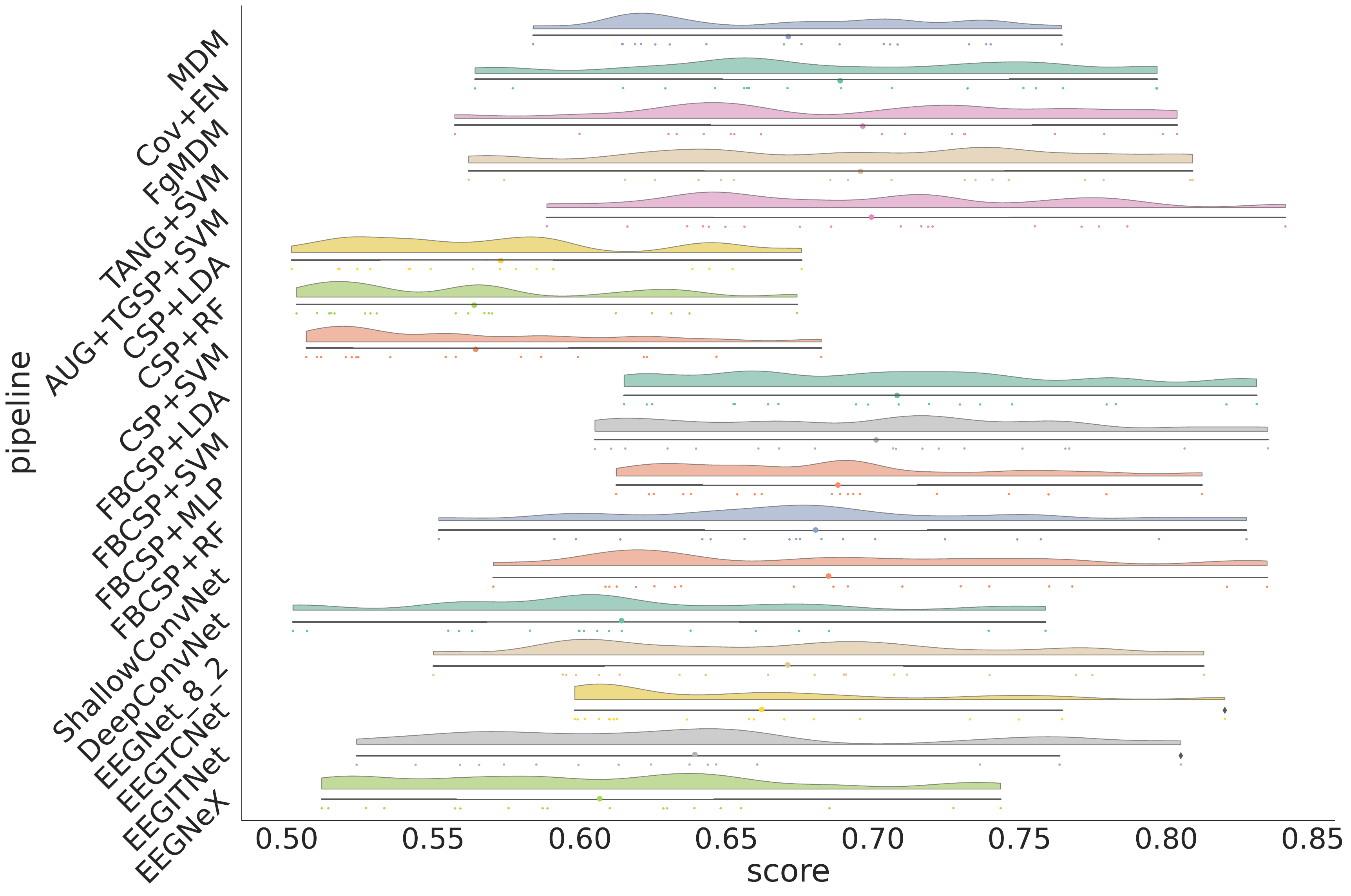}}
             \hfill
     \subfloat[]{%
            \includegraphics[width=0.45\linewidth]{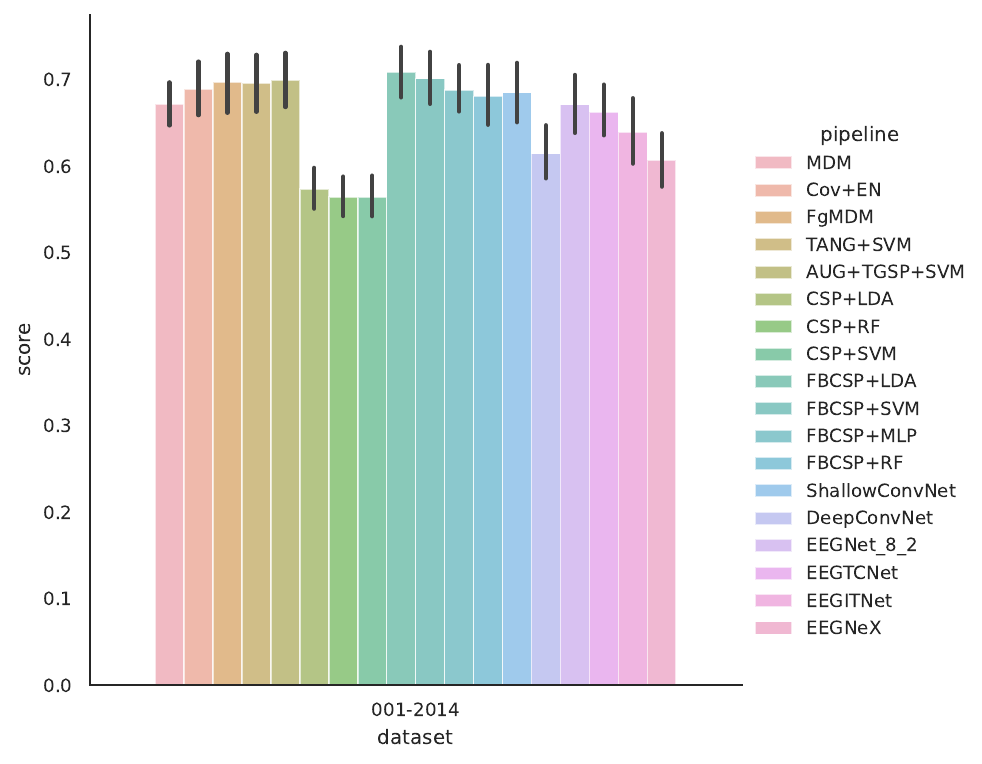}}
     \\
    \subfloat[]{%
        \includegraphics[width=0.6\linewidth]{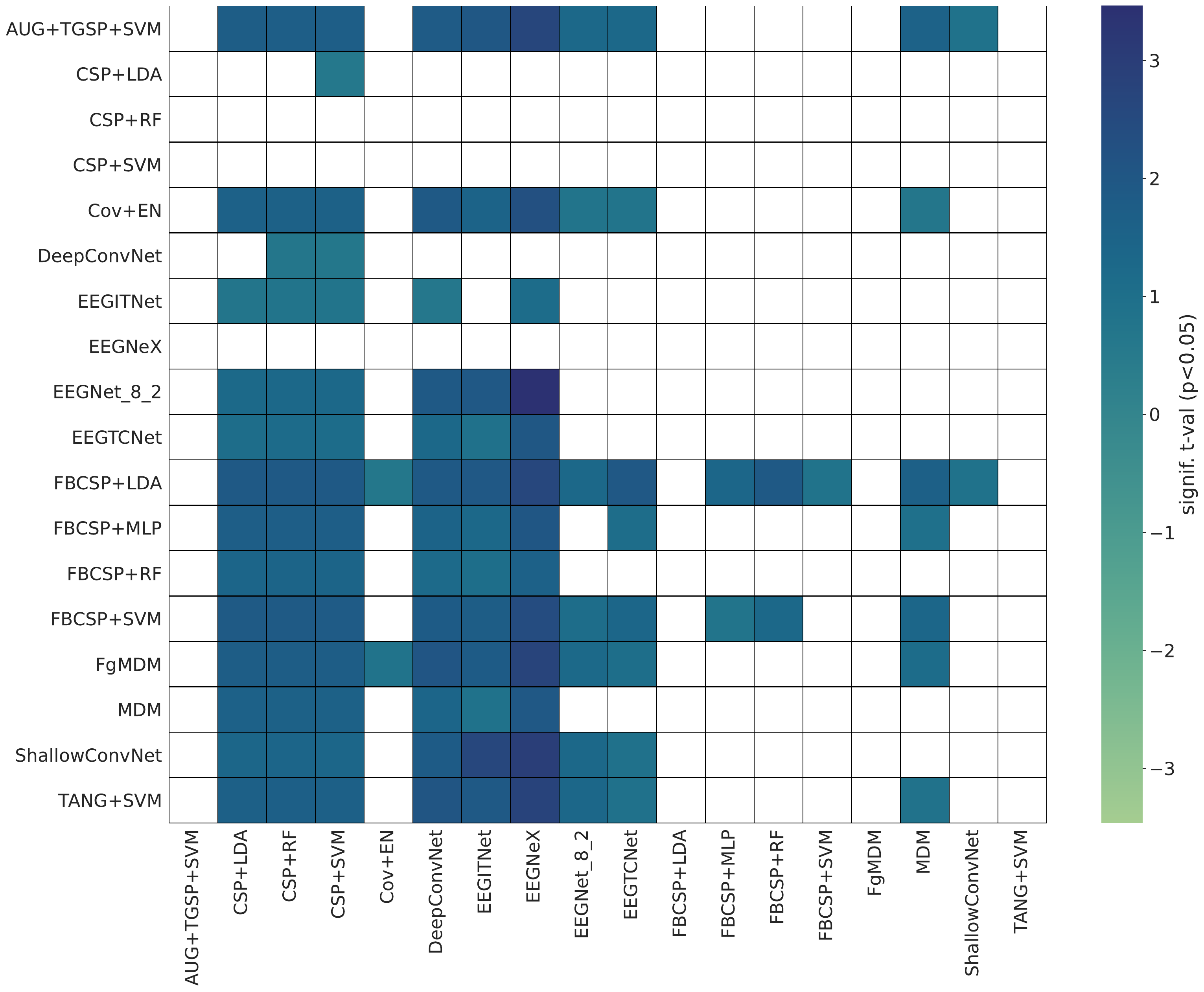}}  
    \caption{Result for BNCI2014001 classification, using Within-Session evaluation. Plot (a) shows the rain clouds plots for each pipeline, showing the distribution of the score of every subject. Plot (b) shows a bar plot of the score with the error of the different pipelines and for every datasets considered. Plot (c) shows the meta analysis of the different methods considered. This plots the significance that the algorithm on the y-axis is better than the one on the x-axis. The color represents the significance level of the difference of accuracy, in terms of t-values, and we show only the significant interactions ($p < 0.05$). 
    }
    \label{fig:BNCI2014001_WS}
\end{figure*}

\newpage

%lhrh_within_MDM
\begin{figure*}[!ht]  
    \centering
    \centering
     \subfloat[]{%
            \includegraphics[width=0.45\linewidth]{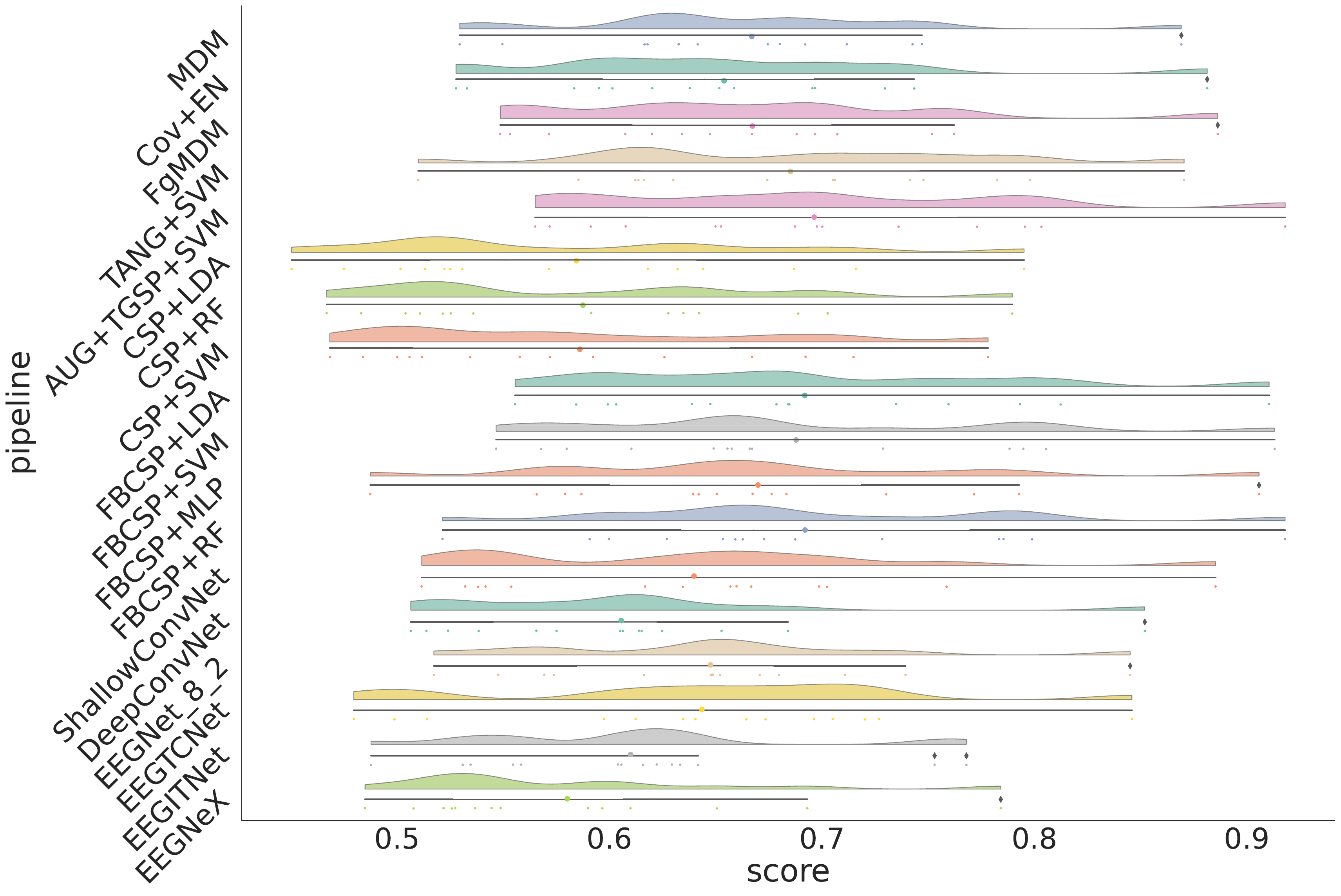}}
             \hfill
     \subfloat[]{%
            \includegraphics[width=0.45\linewidth]{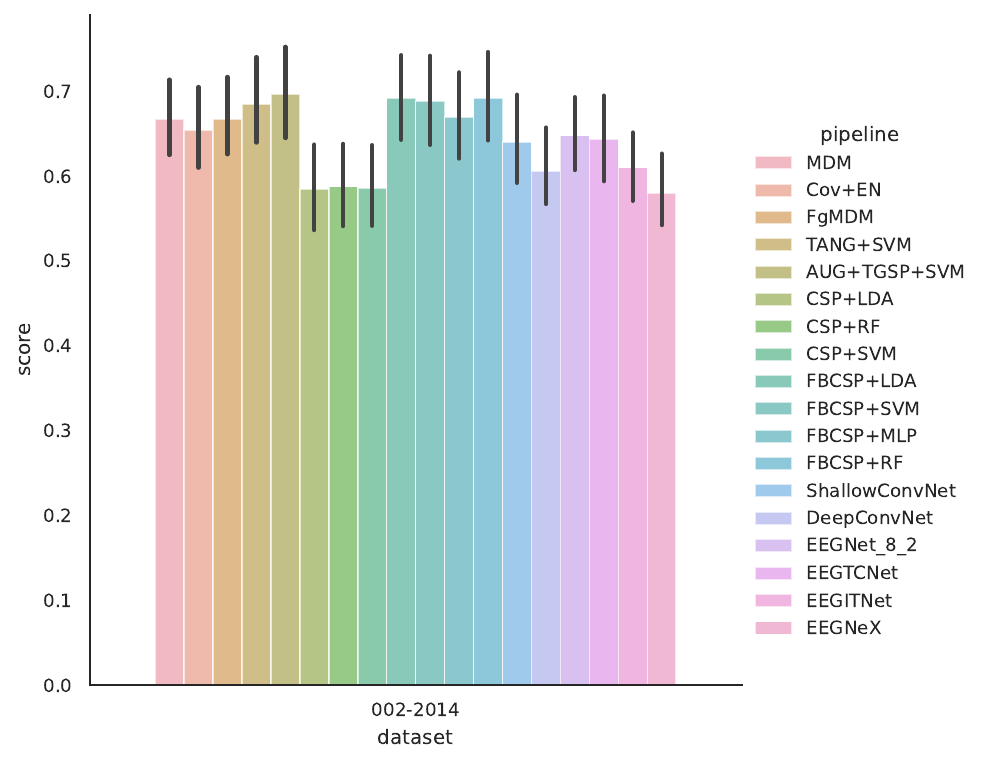}}
     \\
    \subfloat[]{%
        \includegraphics[width=0.6\linewidth]{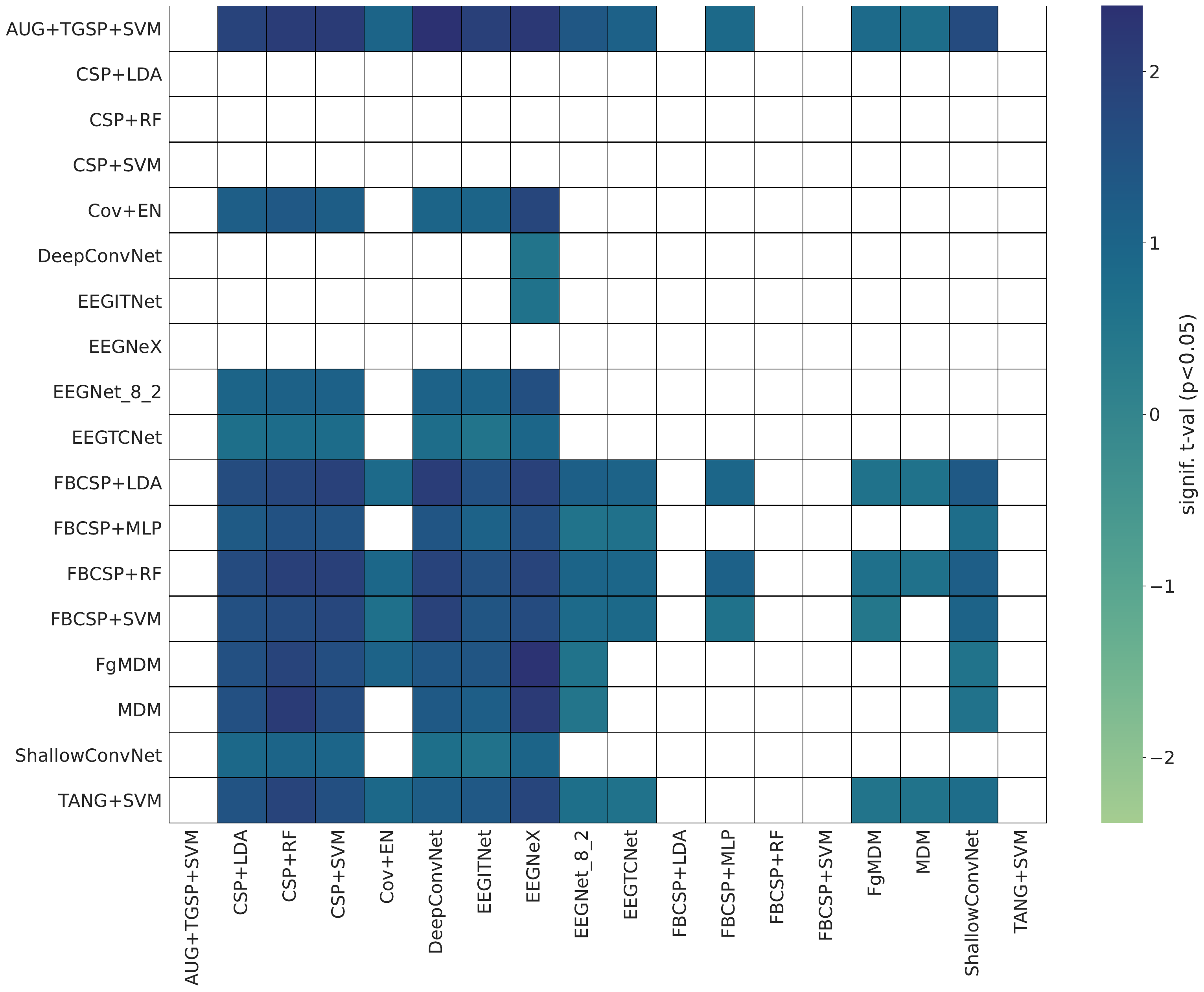}}     
    \caption{Result for BNCI2014002 classification, using Within-Session evaluation. Plot (a) shows the rain clouds plots for each pipeline, showing the distribution of the score of every subject. Plot (b) shows a bar plot of the score with the error of the different pipelines and for every datasets considered. Plot (c) shows the meta analysis of the different methods considered. This plots the significance that the algorithm on the y-axis is better than the one on the x-axis. The color represents the significance level of the difference of accuracy, in terms of t-values, and we show only the significant interactions ($p < 0.05$). 
    }
    \label{fig:BNCI2014002_WS}
\end{figure*}

\newpage

%lhrh_within_MDM
\begin{figure*}[ht]  
    \centering
    \centering
     \subfloat[]{%
            \includegraphics[width=0.45\linewidth]{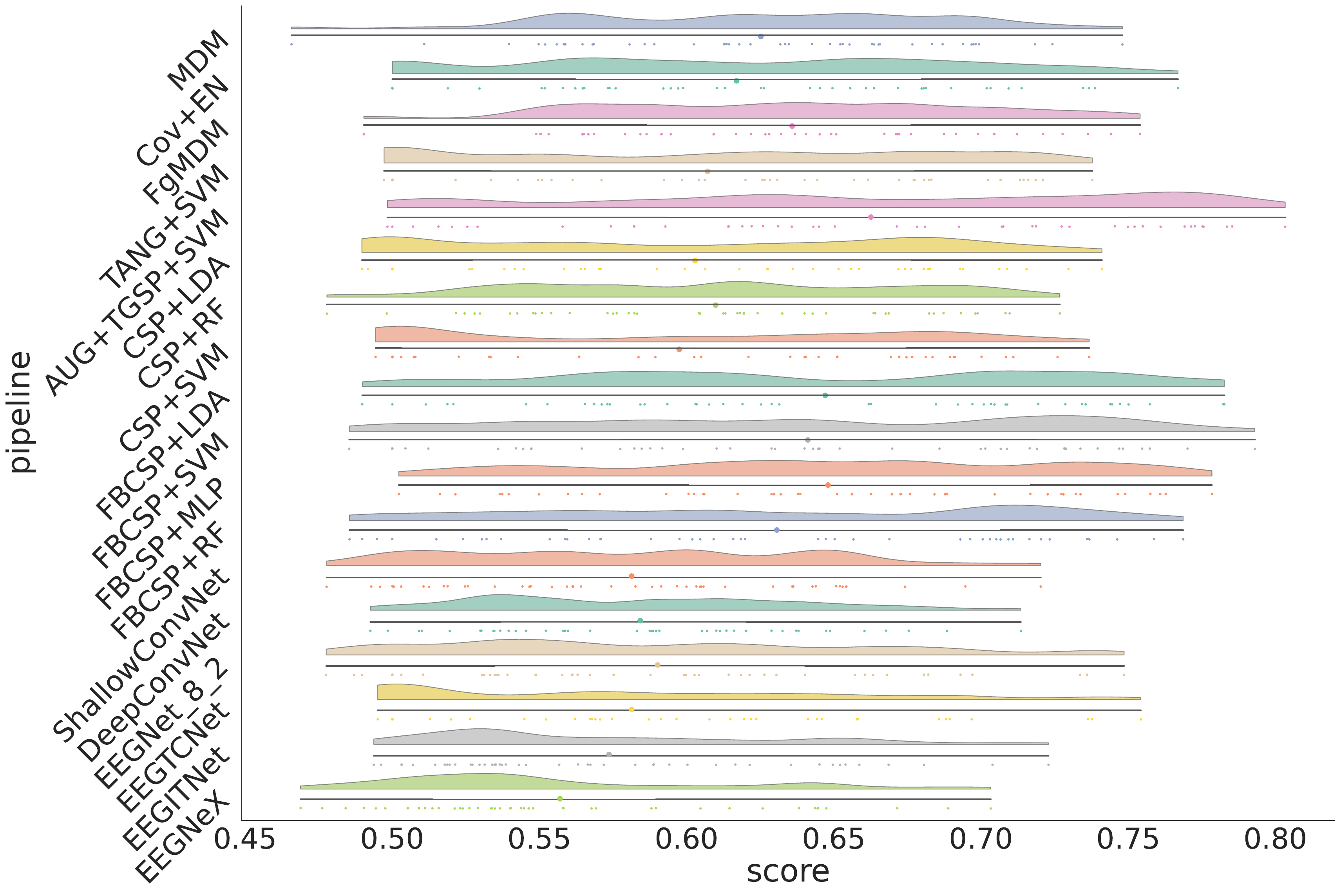}}
             \hfill
     \subfloat[]{%
            \includegraphics[width=0.45\linewidth]{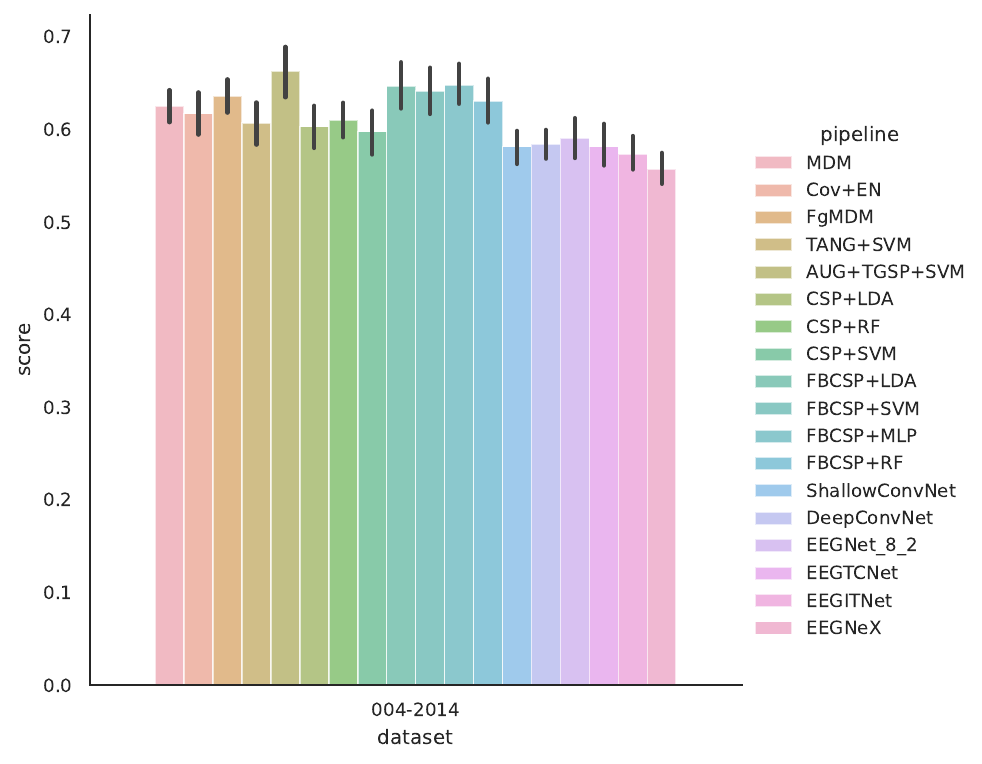}}
     \\
    \subfloat[]{%
        \includegraphics[width=0.6\linewidth]{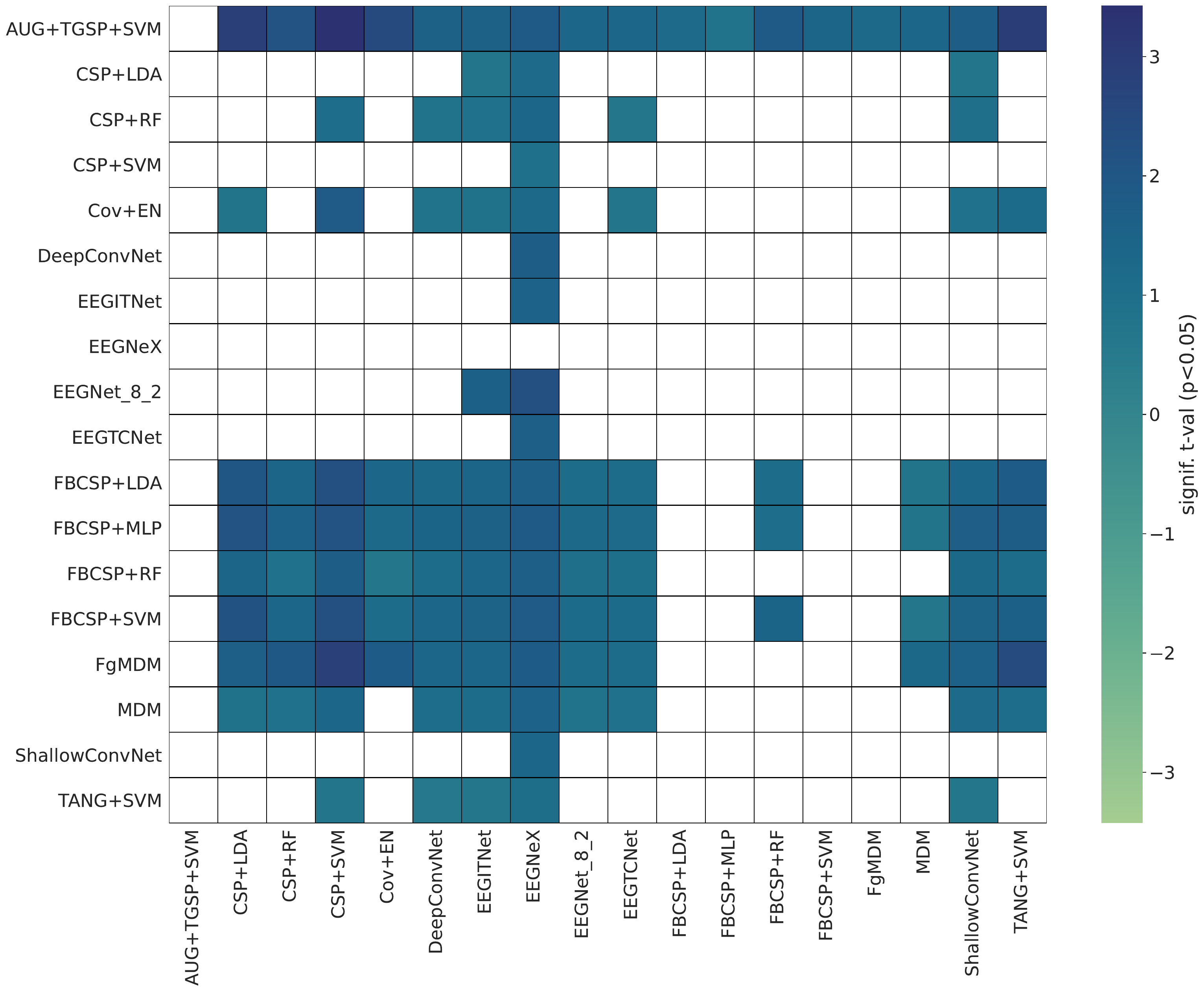}}  
    \caption{Result for BNCI2014004 classification, using Within-Session evaluation. Plot (a) shows the rain clouds plots for each pipeline, showing the distribution of the score of every subject. Plot (b) shows a bar plot of the score with the error of the different pipelines and for every datasets considered. Plot (c) shows the meta analysis of the different methods considered. This plots the significance that the algorithm on the y-axis is better than the one on the x-axis. The color represents the significance level of the difference of accuracy, in terms of t-values, and we show only the significant interactions ($p < 0.05$). 
    }
    \label{fig:BNCI2014004_WS}
\end{figure*}

\newpage

%lhrh_within_MDM
\begin{figure*}[ht]  
    \centering
    \centering
     \subfloat[]{%
            \includegraphics[width=0.45\linewidth]{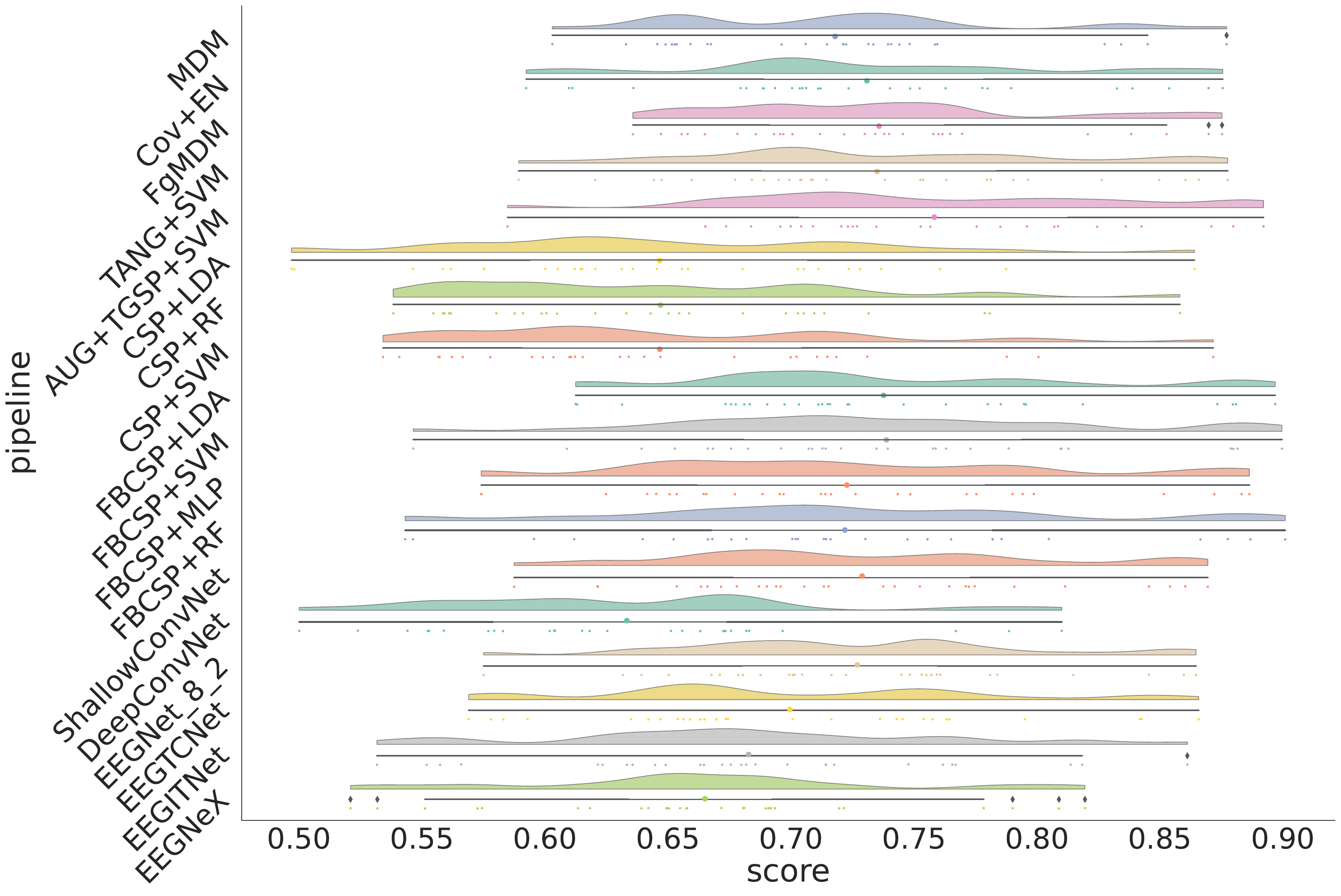}}
             \hfill
     \subfloat[]{%
            \includegraphics[width=0.45\linewidth]{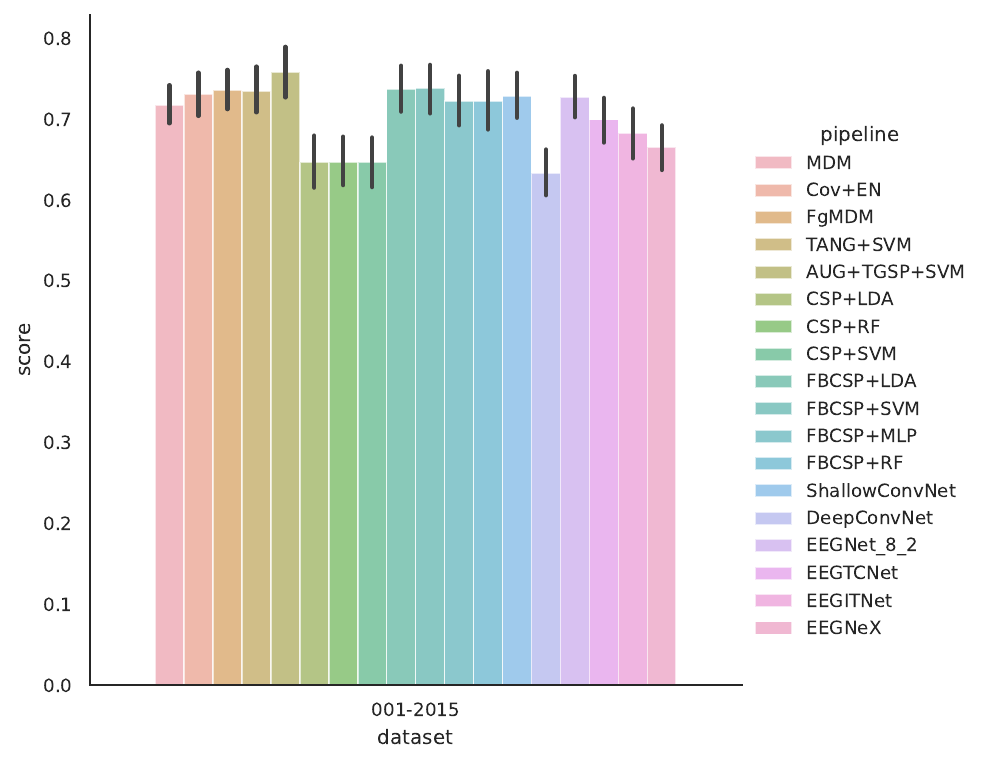}}
     \\
    \subfloat[]{%
        \includegraphics[width=0.6\linewidth]{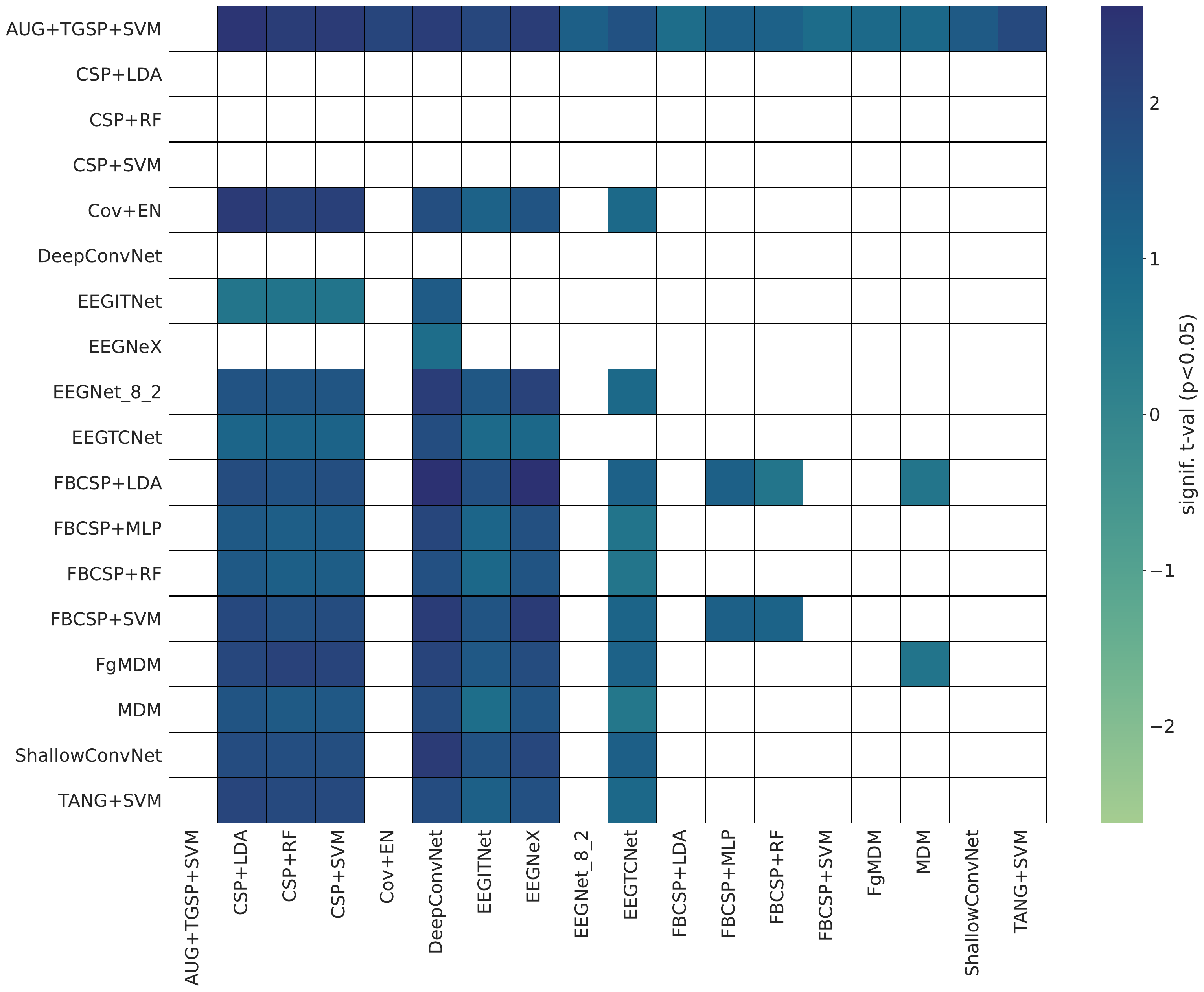}}  
    \caption{Result for BNCI2015001 classification, using Within-Session evaluation. Plot (a) shows the rain clouds plots for each pipeline, showing the distribution of the score of every subject. Plot (b) shows a bar plot of the score with the error of the different pipelines and for every datasets considered. Plot (c) shows the meta analysis of the different methods considered. This plots the significance that the algorithm on the y-axis is better than the one on the x-axis. The color represents the significance level of the difference of accuracy, in terms of t-values, and we show only the significant interactions ($p < 0.05$). 
    }
    \label{fig:BNCI2015001_WS}
\end{figure*}

\begin{figure*}[!ht]  
    \centering
    \centering
     \subfloat[]{%
            \includegraphics[width=0.45\linewidth]{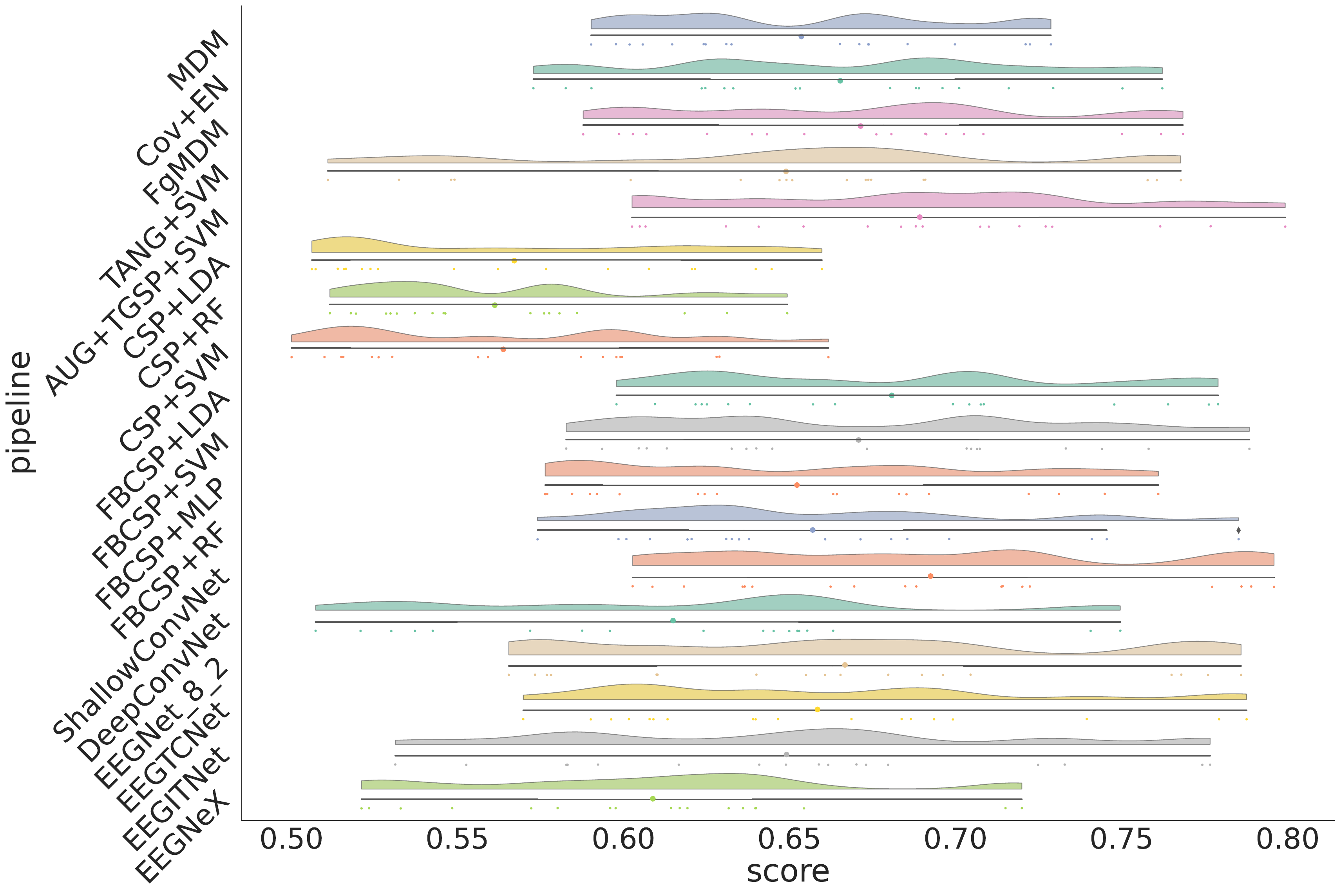}}
             \hfill
     \subfloat[]{%
            \includegraphics[width=0.45\linewidth]{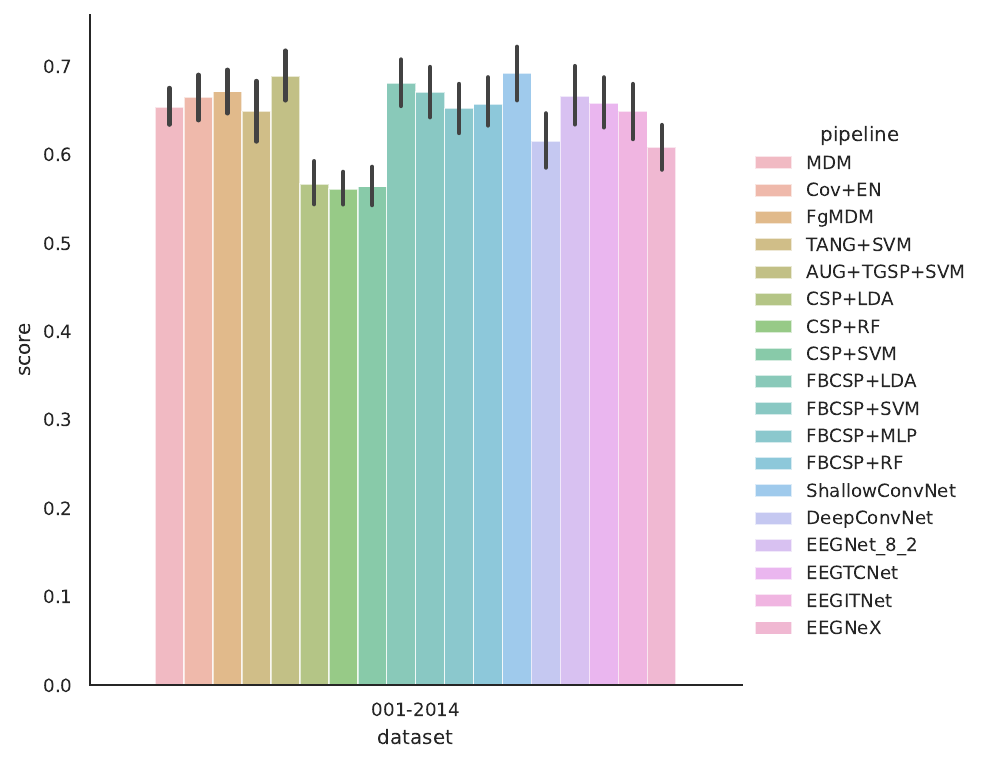}}
     \\
    \subfloat[]{%
        \includegraphics[width=0.6\linewidth]{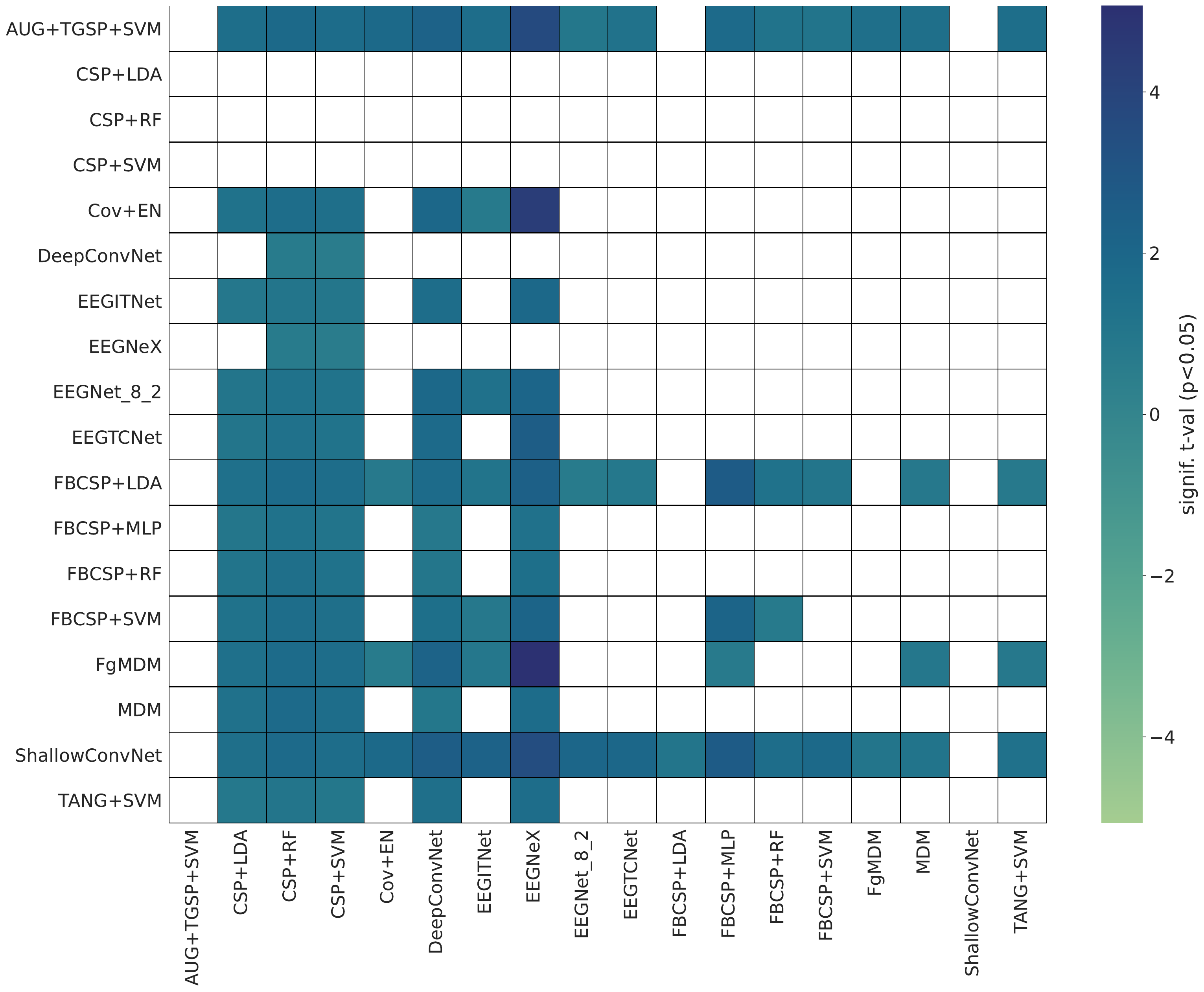}}  
    \caption{Result for BNCI2014001 classification, using Cross-Session evaluation. Plot (a) shows the rain clouds plots for each pipeline, showing the distribution of the score of every subject. Plot (b) shows a bar plot of the score with the error of the different pipelines and for every datasets considered. Plot (c) shows the meta analysis of the different methods considered. This plots the significance that the algorithm on the y-axis is better than the one on the x-axis. The color represents the significance level of the difference of accuracy, in terms of t-values, and we show only the significant interactions ($p < 0.05$). 
    }
    \label{fig:BNCI2014001_CS}
\end{figure*}

\newpage

%lhrh_within_MDM
\begin{figure*}[ht]  
    \centering
    \centering
     \subfloat[]{%
            \includegraphics[width=0.45\linewidth]{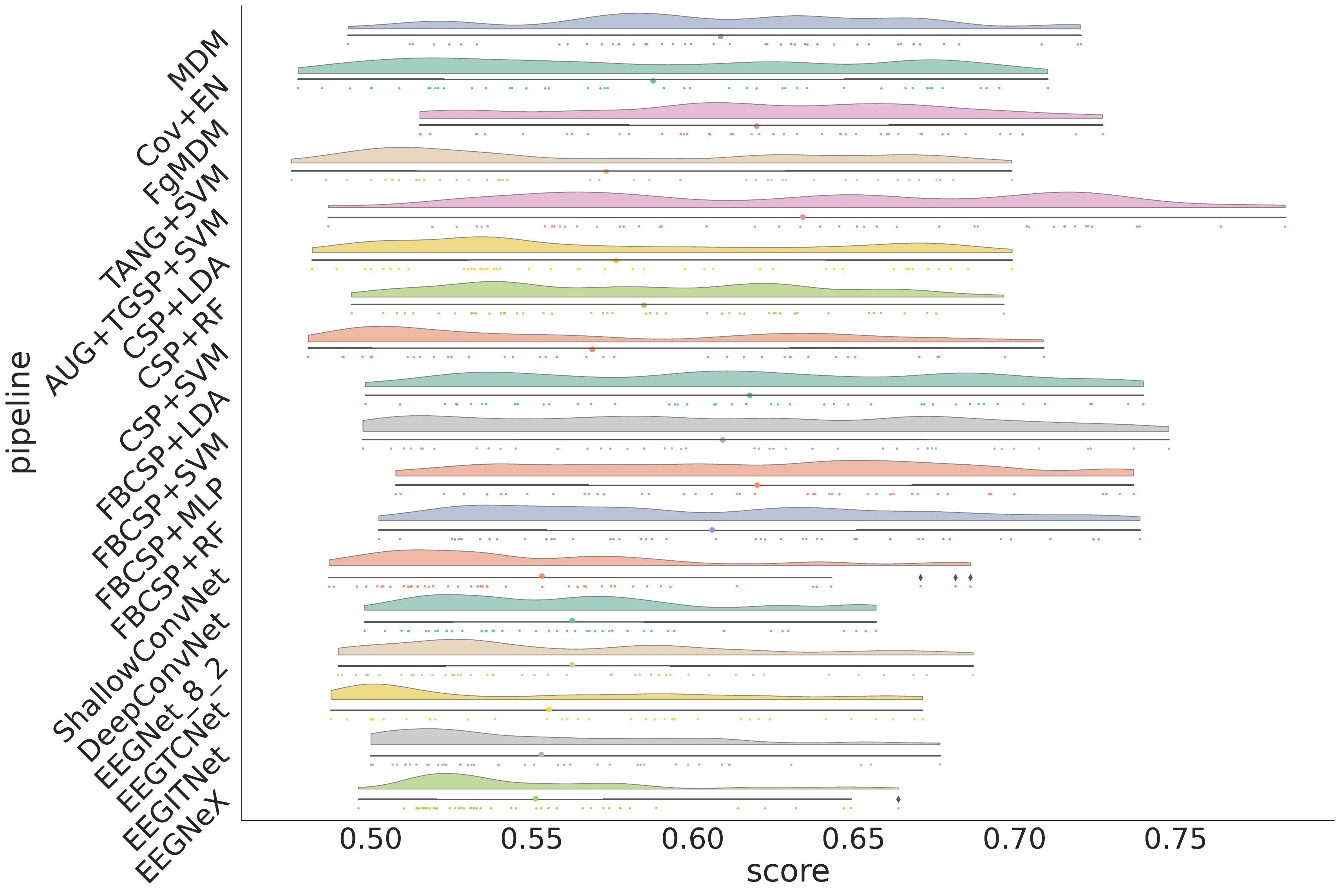}}
             \hfill
     \subfloat[]{%
            \includegraphics[width=0.45\linewidth]{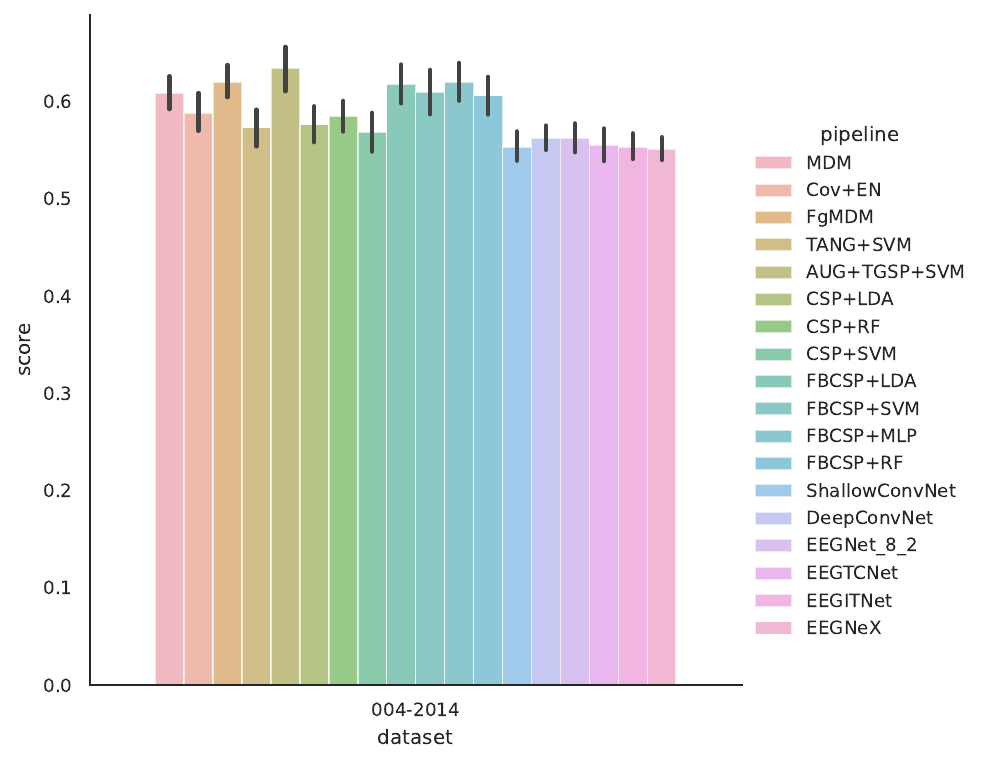}}
     \\
    \subfloat[]{%
        \includegraphics[width=0.6\linewidth]{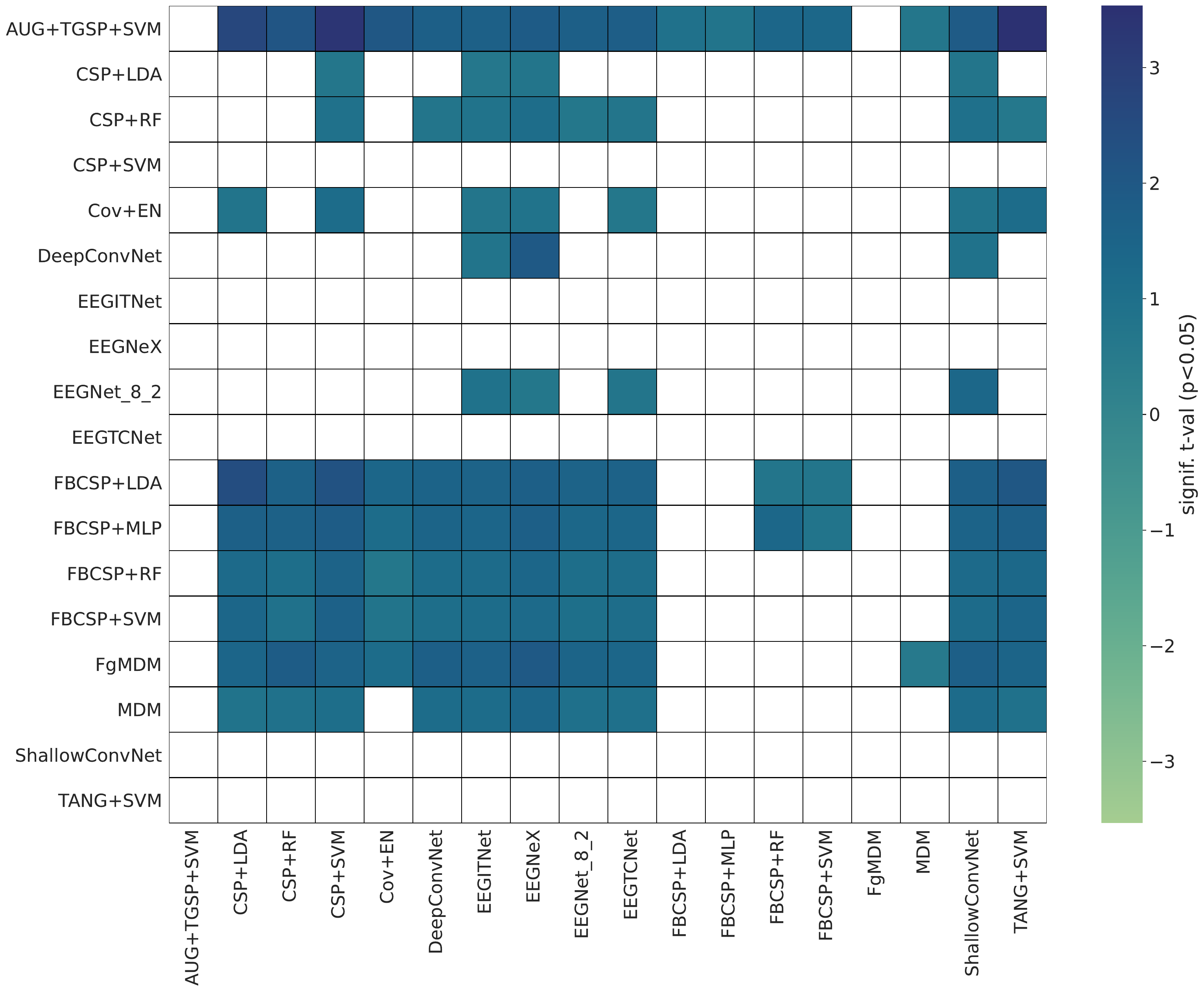}}  
    \caption{Result for BNCI2014004 classification, using Cross-Session evaluation. Plot (a) shows the rain clouds plots for each pipeline, showing the distribution of the score of every subject. Plot (b) shows a bar plot of the score with the error of the different pipelines and for every datasets considered. Plot (c) shows the meta analysis of the different methods considered. This plots the significance that the algorithm on the y-axis is better than the one on the x-axis. The color represents the significance level of the difference of accuracy, in terms of t-values, and we show only the significant interactions ($p < 0.05$). 
    }
    \label{fig:BNCI2014004_CS}
\end{figure*}

\newpage

%lhrh_within_MDM
\begin{figure*}[ht]  
    \centering
    \centering
     \subfloat[]{%
            \includegraphics[width=0.45\linewidth]{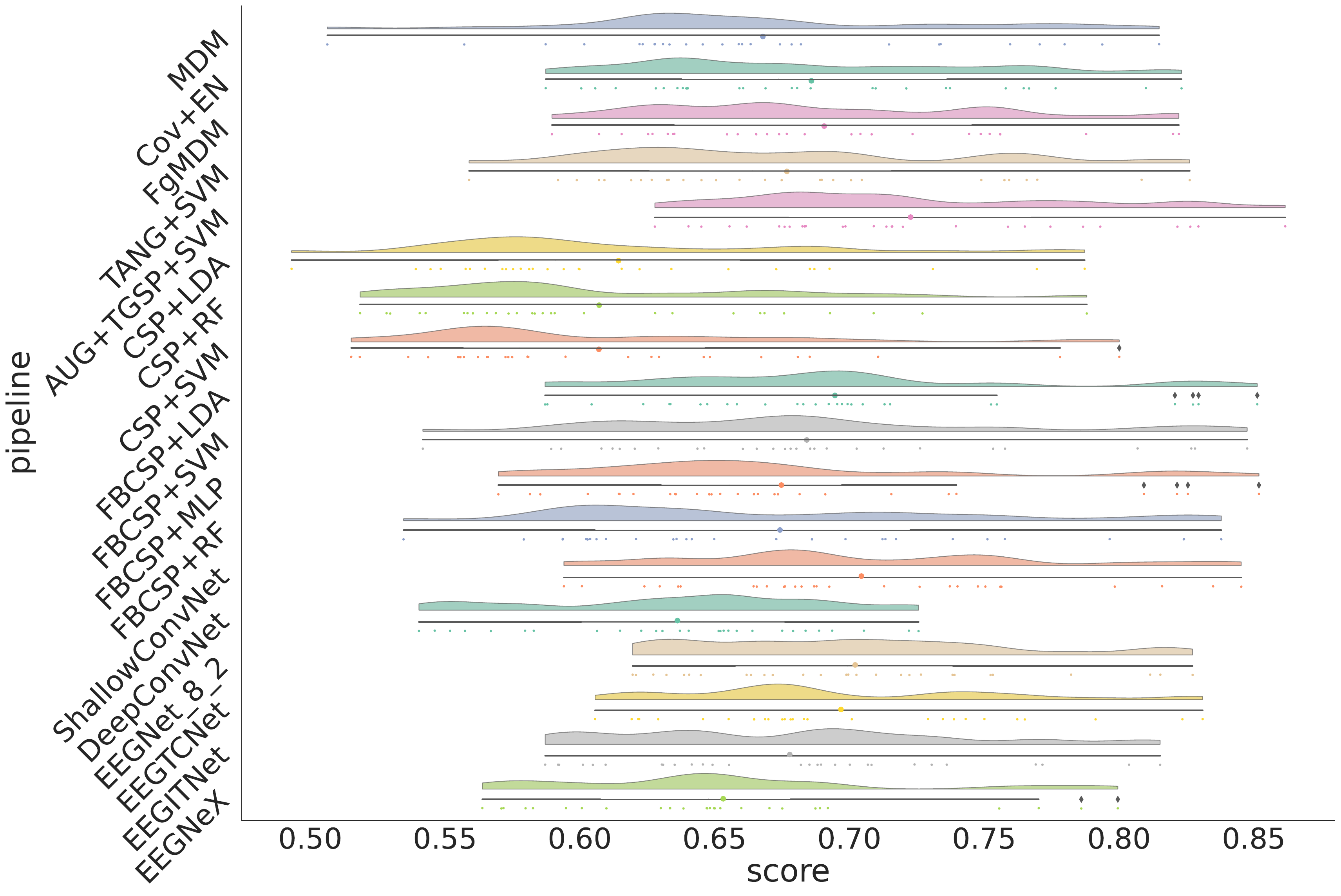}}
             \hfill
     \subfloat[]{%
            \includegraphics[width=0.45\linewidth]{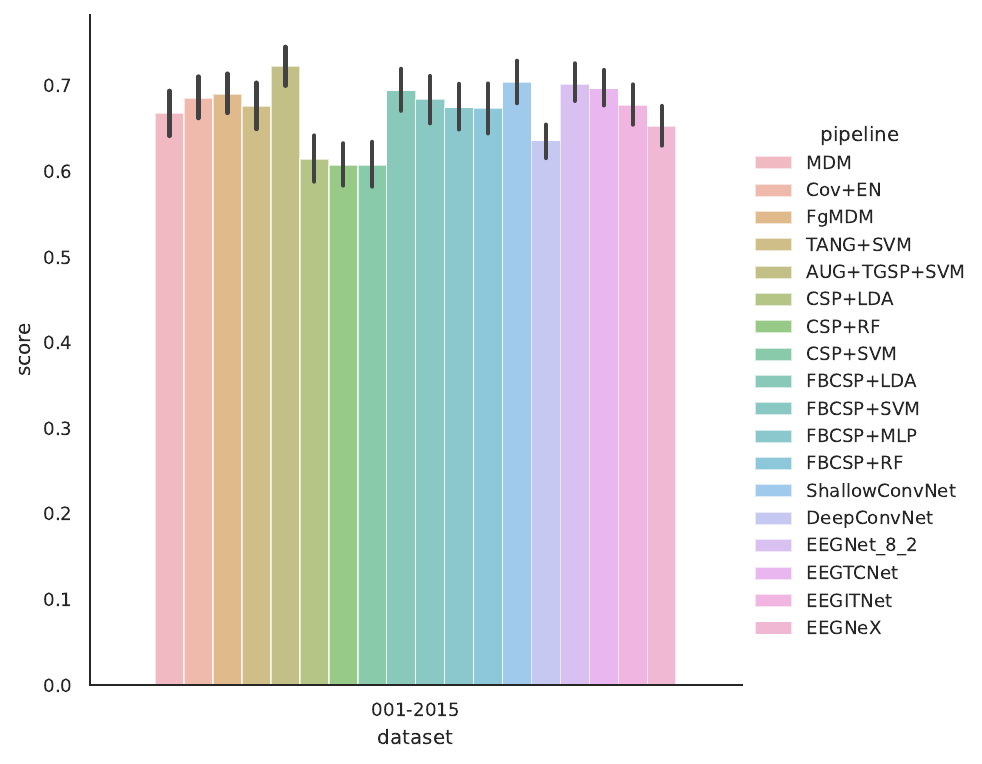}}
     \\
    \subfloat[]{%
        \includegraphics[width=0.6\linewidth]{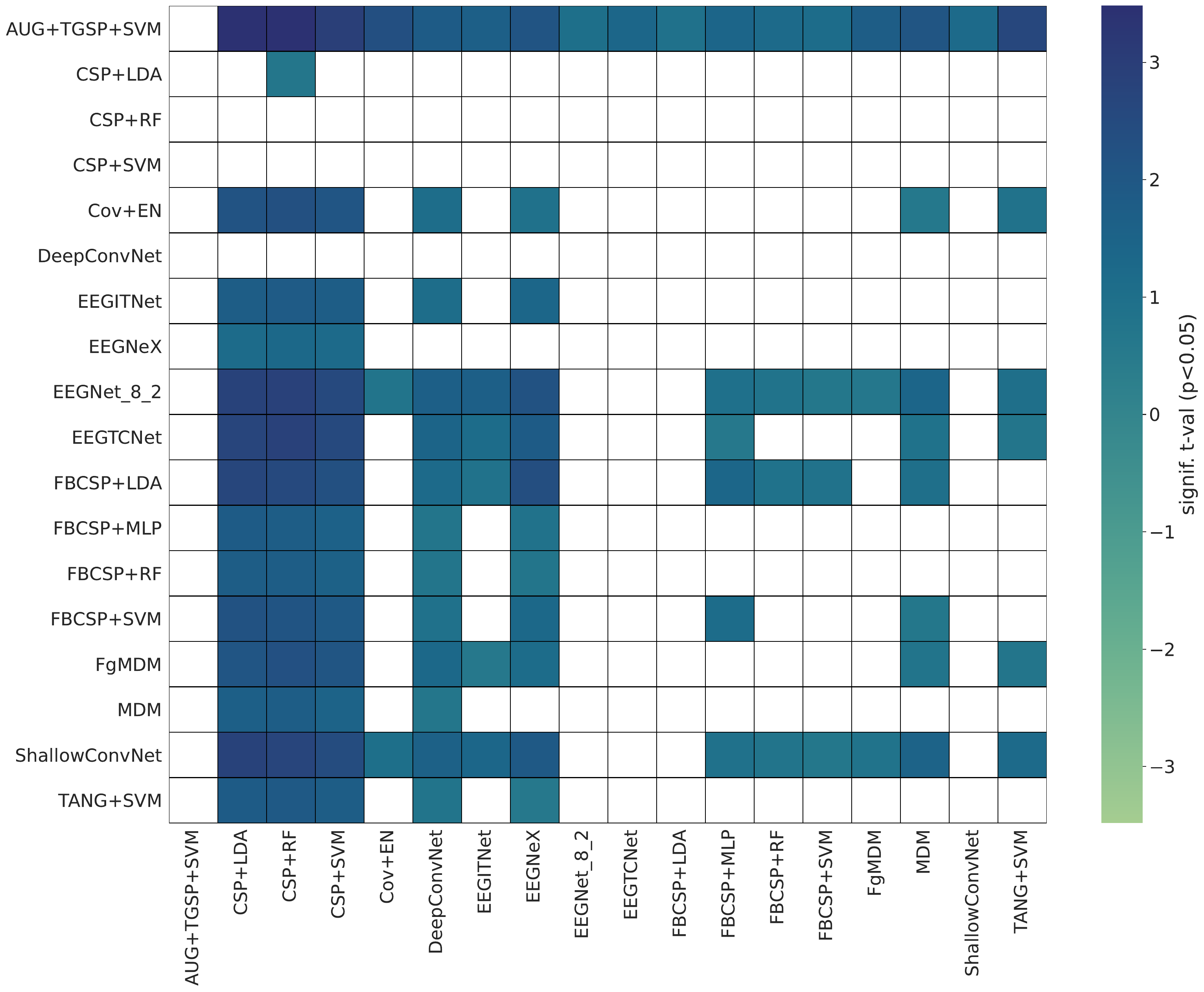}}  
    \caption{Result for BNCI2015001 classification, using Cross-Session evaluation. Plot (a) shows the rain clouds plots for each pipeline, showing the distribution of the score of every subject. Plot (b) shows a bar plot of the score with the error of the different pipelines and for every datasets considered. Plot (c) shows the meta analysis of the different methods considered. This plots the significance that the algorithm on the y-axis is better than the one on the x-axis. The color represents the significance level of the difference of accuracy, in terms of t-values, and we show only the significant interactions ($p < 0.05$). 
    }
    \label{fig:BNCI2015001_CS}
\end{figure*}

\begin{table*}[!ht]
\caption{Performance Offline Within-Session Evaluation using the package MOABB changing the scoring to use nMCC.  Results for the DL architecture are listed after the two line.}
\label{table:Within_Offline}
\centering
\resizebox{0.8\linewidth}{!}{\begin{tabular}{c|c|c|c|c}
Pipeline &                    BNCI2014002    & BNCI2014004          & BNCI2015001      & BNCI2014001\\
\hline MDM &              $0.73 \pm 0.15$ & 0.73 $\pm$ 0.14 &      0.81 $\pm$ 0.14 & $0.81 \pm 0.11$  \\
\hline Cov + EN &         $0.82 \pm 0.12$ & 0.75 $\pm$ 0.14 &      0.86 $\pm$ 0.10 & $0.83 \pm 0.10$  \\
\hline FgMDM &            $0.81 \pm 0.11$ & 0.74 $\pm$ 0.14 &      0.85 $\pm$ 0.11 & $0.81 \pm 0.11$  \\
\hline TANG + SVM &       $0.81 \pm 0.12$ & 0.75 $\pm$ 0.14 &      0.85 $\pm$ 0.11 & $0.82 \pm 0.10$  \\
\hline AUG + TANG + SVM & $0.84 \pm 0.11$ & \textbf{0.81} $\pm$ \textbf{0.12} &      0.90 $\pm$ 0.08 & \textbf{0.86} $\pm$ \textbf{0.09}  \\
\hline CSP + LDA &        $0.80 \pm 0.13$ & 0.74 $\pm$ 0.14 &      0.84 $\pm$ 0.11 & $0.79 \pm 0.10$  \\
\hline CSP + RF &         $0.79 \pm 0.12$ & 0.72 $\pm$ 0.14 &      0.83 $\pm$ 0.11 & $0.78 \pm 0.10$  \\
\hline CSP + SVM &        $0.80 \pm 0.13$ & 0.75 $\pm$ 0.15 &      0.84 $\pm$ 0.10 & $0.80 \pm 0.11$  \\
\hline FBCSP+LDA &        $0.81 \pm 0.13$ & 0.77 $\pm$ 0.14 &      0.88 $\pm$ 0.09 & $0.84 \pm 0.09$  \\
\hline FBCSP+SVM &        $0.82 \pm 0.12$ & 0.78 $\pm$ 0.13 &      0.88 $\pm$ 0.08 & $0.84 \pm 0.09$  \\
\hline FBCSP+MLP &        $0.81 \pm 0.12$ & 0.78 $\pm$ 0.14 &      0.87 $\pm$ 0.09 & $0.83 \pm 0.09$  \\
\hline FBCSP+RF &         $0.81 \pm 0.13$ & 0.75 $\pm$ 0.14 &      0.86 $\pm$ 0.09 & $0.82 \pm 0.09$  \\ \hline
\hline ShallowConvNet &   $\textbf{0.88} \pm \textbf{0.12}$ & 0.72 $\pm$ 0.18 &      \textbf{0.91} $\pm$ \textbf{0.11} & $0.72 \pm 0.17$  \\
\hline DeepConvNet &      $0.87 \pm 0.11$ & 0.72 $\pm$ 0.19 &      0.88 $\pm$ 0.14 & $0.34 \pm 0.08$  \\
\hline EEGNet 8 2 &       $0.85 \pm 0.16$ & 0.69 $\pm$ 0.20 &      0.90 $\pm$ 0.12 & $0.61 \pm 0.21$  \\
\hline EEG ITNet &        $0.70 \pm 0.18$ & 0.65 $\pm$ 0.15 &      0.71 $\pm$ 0.17 & $0.34 \pm 0.05$  \\
\hline EEG TCNet &        $0.73 \pm 0.20$ & 0.69 $\pm$ 0.20 &      0.76 $\pm$ 0.19 & $0.40 \pm 0.14$  \\
\hline EEGNeX 8 32 &      $0.70 \pm 0.21$ & 0.67 $\pm$ 0.17 &      0.72 $\pm$ 0.20 & $0.45 \pm 0.16$  \\
\hline 
\end{tabular}
}
\end{table*}

\begin{table*}[!ht]
\caption{Performance Offline Cross-Session Evaluation using the package MOABB changing the scoring to use nMCC. Results for the DL architecture are listed after the two line.}
\label{table:Cross_Offline}
\centering
\resizebox{0.65\linewidth}{!}{\begin{tabular}{c|c|c|c}
Pipeline &                  BNCI2014004          & BNCI2015001      & BNCI2014001\\
\hline MDM &              0.79 $\pm$ 0.14 &      0.87 $\pm$ 0.11 & $0.59 \pm 0.14$  \\
\hline Cov + EN &         0.81 $\pm$ 0.14 &      0.90 $\pm$ 0.10 & $0.64 \pm 0.12$  \\
\hline FgMDM &            0.80 $\pm$ 0.14 &      0.89 $\pm$ 0.10 & $0.63 \pm 0.13$  \\
\hline TANG + SVM &       0.81 $\pm$ 0.14 &      0.90 $\pm$ 0.10 & $0.62 \pm 0.13$  \\
\hline AUG + TANG + SVM & \textbf{0.85} $\pm$ \textbf{0.14} &      \textbf{0.94} $\pm$ \textbf{0.07} & \textbf{0.73} $\pm$ \textbf{0.13}  \\
\hline CSP + LDA &        0.81 $\pm$ 0.14 &      0.89 $\pm$ 0.10 & $0.60 \pm 0.14$  \\
\hline CSP + RF &         0.76 $\pm$ 0.15 &      0.86 $\pm$ 0.12 & $0.56 \pm 0.13$  \\
\hline CSP + SVM &        0.81 $\pm$ 0.14 &      0.89 $\pm$ 0.10 & $0.61 \pm 0.13$  \\
\hline FBCSP+LDA &        0.82 $\pm$ 0.14 &      0.91 $\pm$ 0.08 & $0.66 \pm 0.13$  \\
\hline FBCSP+SVM &        0.83 $\pm$ 0.14 &      0.91 $\pm$ 0.08 & $0.66 \pm 0.13$  \\
\hline FBCSP+MLP &        0.82 $\pm$ 0.14 &      0.92 $\pm$ 0.08 & $0.65 \pm 0.12$  \\
\hline FBCSP+RF &         0.80 $\pm$ 0.15 &      0.90 $\pm$ 0.09 & $0.63 \pm 0.11$  \\ \hline
\hline ShallowConvNet &   0.73 $\pm$ 0.19 &      0.92 $\pm$ 0.10 & $0.70 \pm 0.16$  \\
\hline DeepConvNet &      0.75 $\pm$ 0.17 &      0.90 $\pm$ 0.11 & $0.37 \pm 0.10$  \\
\hline EEGNet 8 2 &       0.75 $\pm$ 0.16 &      0.90 $\pm$ 0.12 & $0.59 \pm 0.19$  \\
\hline EEG ITNet &        0.71 $\pm$ 0.15 &      0.79 $\pm$ 0.15 & $0.43 \pm 0.16$  \\
\hline EEG TCNet &        0.74 $\pm$ 0.20 &      0.84 $\pm$ 0.17 & $0.44 \pm 0.14$  \\
\hline EEGNeX 8 32 &      0.71 $\pm$ 0.16 &      0.76 $\pm$ 0.19 & $0.46 \pm 0.16$  \\
\hline 
\end{tabular}
}
\end{table*}

\begin{table*}[!ht]
\caption{Performance Pseudo Online Cross-Session Evaluation. Results for the DL architecture are listed after the two line.}
\label{table:Cross_Flat}
\centering
\resizebox{0.65\linewidth}{!}{\begin{tabular}{c|c|c|c}
Pipeline &                BNCI2014004           & BNCI2015001       & BNCI2014001\\
\hline MDM &              0.61 $\pm$ 0.06 &      0.67 $\pm$ 0.07 & $0.65 \pm 0.05$  \\
\hline Cov + EN &         0.59 $\pm$ 0.07 &      0.69 $\pm$ 0.07 & $0.67 \pm 0.06$  \\
\hline FgMDM &            0.62 $\pm$ 0.06 &      0.69 $\pm$ 0.06 & $0.67 \pm 0.06$  \\
\hline TANG + SVM &       0.58 $\pm$ 0.07 &      0.68 $\pm$ 0.07 & $0.66 \pm 0.06$  \\
\hline AUG + TANG + SVM & \textbf{0.64} $\pm$ \textbf{0.08} &      \textbf{0.73} $\pm$ \textbf{0.06} & $\textbf{0.70} \pm \textbf{0.06}$  \\
\hline CSP + LDA &        0.58 $\pm$ 0.07 &      0.62 $\pm$ 0.07 & $0.57 \pm 0.05$  \\
\hline CSP + RF &         0.58 $\pm$ 0.06 &      0.61 $\pm$ 0.07 & $0.57 \pm 0.04$  \\
\hline CSP + SVM &        0.60 $\pm$ 0.07 &      0.61 $\pm$ 0.07 & $0.57 \pm 0.05$  \\
\hline FBCSP+LDA &        0.62 $\pm$ 0.07 &      0.70 $\pm$ 0.07 & $0.68 \pm 0.06$  \\
\hline FBCSP+SVM &        0.61 $\pm$ 0.07 &      0.70 $\pm$ 0.08 & $0.68 \pm 0.05$  \\
\hline FBCSP+MLP &        0.63 $\pm$ 0.07 &      0.69 $\pm$ 0.08 & $0.68 \pm 0.06$  \\
\hline FBCSP+RF &         0.61 $\pm$ 0.07 &      0.68 $\pm$ 0.08 & $0.66 \pm 0.06$  \\ \hline
\hline ShallowConvNet &   0.55 $\pm$ 0.05 &      0.71 $\pm$ 0.07 & $0.69 \pm 0.06$  \\
\hline DeepConvNet &      0.56 $\pm$ 0.05 &      0.64 $\pm$ 0.06 & $0.62 \pm 0.07$  \\
\hline EEGNet 8 2 &       0.56 $\pm$ 0.06 &      0.70 $\pm$ 0.07 & $0.67 \pm 0.08$  \\
\hline EEG ITNet &        0.55 $\pm$ 0.05 &      0.68 $\pm$ 0.07 & $0.64 \pm 0.07$  \\
\hline EEG TCNet &        0.55 $\pm$ 0.06 &      0.70 $\pm$ 0.07 & $0.66 \pm 0.06$  \\
\hline EEGNeX 8 32 &      0.56 $\pm$ 0.05 &      0.66 $\pm$ 0.07 & $0.61 \pm 0.06$  \\
\hline 
\end{tabular}
}
\end{table*}

\end{document}